\documentclass[draft]{agujournal2019}
\usepackage{url} %this package should fix any errors with URLs in refs.
\usepackage[utf8]{inputenc}
\usepackage{lineno}
\usepackage[inline]{trackchanges} 
\usepackage{natbib}
\usepackage{soul}
\usepackage{graphicx}
\usepackage{multirow}
\usepackage{helvet}
\usepackage{amsmath}
\usepackage{color}
\usepackage{colortbl}
\usepackage{setspace}
\usepackage{outlines}
\usepackage{cancel}
\usepackage{todonotes}
\draftfalse
\graphicspath{{Figures_low/}}
\journalname{JGR: Space Physics}
\newcommand{\planss}[1]{Plan. Sp. Sci.}
\newcommand{\icarus}[1]{Icarus}

\addeditor{sb}
\addeditor{gg}
\addeditor{sh}
\newcolumntype{a}{>{\columncolor{green!15}}c}
\newcolumntype{q}{>{\columncolor{blue!15}}c}
\newcolumntype{s}{>{\columncolor{red!15}}c}
\begin{document}
\begin{singlespace}
	\title{The Sun-Earth-Moon Connection:\\ II--Solar Wind and Lunar Surface Interaction}
	\authors{Suleiman M Baraka \affil{1,6}, Sona Hosseini \affil{2},  Guillaume Gronoff\affil{3,7}, Lotfi Ben$-$Jaffel\affil{4}, Robert Rankin\affil{5}}
	\affiliation{1}{National Institute of Aerospace, 100 Exploration Way, Hampton, VA 23666}
	\affiliation{2}{Jet Propulsion Laboratory, California Institute of Technology, M/S 183–401, 4800 Oak Grove Drive, Pasadena, CA 91109, USA}
	\affiliation{3}{Chemistry and Dynamics Branch, Science Directorate, NASA Langley Research Center, Hampton, Virginia, USA}
\affiliation{4}{Institut d’Astrophysique de Paris, UMR7095, Sorbonne Université, Paris, France}
	\affiliation{5}{Department of Physics, University of Alberta, Edmonton, Alberta, Canada,}
	\affiliation{6}{Department of Physics and Astronomy, University of Calgary, Calgary, Alberta, Canada.}	
 	\affiliation{7}{Science Systems and Application Inc, Hampton, Va, USA}
  \correspondingauthor{Suleiman M Baraka}{suleiman.baraka@nianet.org}
	\begin{keypoints}
	\item A fully kinetic simulation of the Sun-Earth-moon system was performed
	\item The Lunar surface charging and the wake dynamics were evaluated
	\item The Earth's magnetotail does not prevent solar-wind ions and ionospheric ions from reaching the Lunar environment.
\end{keypoints}
\end{singlespace}
\begin{abstract}
{\bf Context:}
	In the pursuit of lunar exploration and the investigation of water presence on the lunar surface, a comprehensive understanding of plasma-surface interactions is crucial since the regolith's space weathering can create H$_2$O. However, the Moon is in the Earth's magnetotail for nearly 20\% of its orbit, which could affect this water creation on the side facing the Earth if this condition shields it from the solar wind.
 
		{\bf Aims:}
    The objective of this study is to understand how the passage of the Moon in the Earth's magnetotail affects the plasma delivery near the lunar surface.
	{\bf Method:}
		The Particle-In-Cell Electromagnetic (EM) Relativistic Global Model, known as IAPIC, is employed to kinetically simulate the Solar Wind-Magnetosphere-Ionosphere-Moon Coupling. 
{\bf Results:}
The Earth's magnetotail does not prevent the influx of solar wind ions and ionospheric ions into the solar environment; therefore the space weathering of the regolith is not stopped in these conditions. In addition, the charge separation of solar wind ions and electrons happens is modeled, leading to electric fields and charging of the lunar surface that can be validated by observations.
		{\bf Conclusion:}
  The study of the Sun-Earth-Moon system provides insight into the lunar environment while in the magnetotail, which is essential to better interpret the results of future Lunar missions. It also provides insights in the Lunar charging in different conditions that could affect the human presence on the Moon.
  \end{abstract}
\clearpage
\section{Introduction}\label{intro}
\justifying
	As exploration efforts in the Lunar environment pick up pace, many questions arise concerning the source of water, its potential movement towards the poles, and its accessibility for human exploration. Among these questions is the extent to which the solar wind (SW) contributes to water production by implanting particles. The Lunar regolith, which comprises a large percentage the surface of the Moon, is formed by a constant barrage of large and small meteorites, as well as the unrelenting impact of solar wind. As a result, it serves as the actual boundary between the Moon and the surrounding interplanetary magnetic field and plasmas  \citep[][and references therein]{Mckay1991}.
 {The investigation of the interaction of solar wind protons with the Lunar regolith using data from the Lunar Prospector mission found that the solar wind protons can penetrate up to several tens of centimeters into the Lunar regolith, causing a significant chemical alteration of the surface materials \citep{Lucey2006}. Multiple investigations have demonstrated that the $\mathrm{H^+}$ ions found in the solar wind have the propensity to react with the upper layers of the Lunar regolith, as well as the rocks present on the surface \citep{Taylor1982,Halekas2012,Poppe2019}}. { When the energetic ions interact with the crystalline structure, they prompt O atoms to respond, which can eventually lead to the creation of OH or $\mathrm{H_2O}$. Recent observations from the Lunar surface by Chang'e 5, as reported by \citet{Mamo2022,Lin2022}, indicate that a significant quantity of water is present in undisturbed regolith. In contrast, the regolith that has been disturbed by the impact of the lander's rockets, i.e., the lower layers, was found to be relatively $\mathrm{H_2O}$-depleted.}   
	{Water molecules are a major component of the Lunar exosphere, resulting from various processes such as chemical reactions and the release of water from meteorite impacts. The studying of the water amount variations in the Lunar exosphere during the Moon's passage through the Earth's magnetotail could help understand the relative importance of these processes \citep[see e.g.][]{Wang2021}. This method assumed that the solar wind cannot directly produce water during these conditions.} 
	{However, while the magnetosphere is a barrier against the direct impact of the solar wind, recent research has revealed that its behavior is more intricate than previously believed \citep{Egan2019}. Recent measurements have indeed been challenging the belief that the magnetosphere was protecting against atmospheric escape.  Notably, polar outflow can trigger the escape of significant amounts of $\mathrm{O^+}$ through the poles, comparable to the rate of escape for unmagnetized planets \citep{Gronoff2020}. As a result, it is crucial to revise our understanding of the Earth's magnetotail surroundings and its impact on the Lunar environment. Previous researches have concentrated on the effect of the passage of the Moon through the Earth's magnetotail on the Lunar exosphere \citep{Wilson2006,Sarantos2008}, showing that it affects its composition and impacts the sodium plume.} \par
{The solar wind can also induce surface charging on the Lunar surface, which can be a significant concern for human activities due to the risk of equipment discharges that may jeopardize their life support and equipment. Earlier studies \citep{Stubbs2007} have addressed the electrostatic charging of the Lunar surface and the electrically driven dust transport.} {According to \citet{Poppe2021a}, surface charging on the Lunar surface is attributed to tail plasma charging and an additional source generated by currents produced by micrometeorite impacts, which contribute 15\%-40\% to the total charging. \citet{Bale1997} discussed the charge imbalance resulting from the topological differences that induce charge fields in the solar wind plasma near the Moon. \citet{Kimura2008} discovered that ions dominate the dayside while electrons dominate the nightside/wake, causing a considerable potential drop.} 
{Lunar Prospector confirmed the theoretically predicted prevalence of negative charges on the shaded side of the Moon \citep{Halekas2005}. Nonetheless, the impact of the magnetotail on surface charging is still unclear.}\par 
{Several techniques have been employed to better understand the Lunar plasma environment and address these questions. For instance, space probes such as WIND or NASA/THEMIS-ARTEMIS have been utilized in extensive research \citep{Halekas2011, Poppe2014, Poppe2018, Poppe2021a, Fatemi2014, Xu2019, Akay2019, Rasca2021} to probe in-situ plasma parameters. The magnetic field within the lunar wake is amplified by a factor of 1.4, and this enhancement is accompanied by the presence of asymmetry. Furthermore, the {MHD} simulation allowed for the calculation of the induced electric field resulting from convection.\citep[e.g.][]{Wang2011,Xie2013}. In addition to space probes, kinetic simulations \citep[e.g.][]{Birch2001} have been extensively utilized to supplement the understanding of the Lunar plasma environment.}\par {The MHD simulation in \citet{Xie2013} provided insights into the interaction between the solar wind and the Moon, with different IMF orientations including north, radial, and tilted. The simulations were compared with NASA/WIND observations in the Lunar wake, revealing the significant role of electron dynamics in the SW-Moon interaction. Moreover, global hybrid simulations have been employed to model the WIND data and the Lunar wake \citep[e.g.][]{Travnivcek2005, Fuqua2019, Jin2020, Rasca2021, Omelchenko2021, Rasca2022}. }\par
In this paper, we present a novel approach to studying the plasma environment of the Sun-Earth-Moon system and its connections using the Particle-In-Cell Electrodynamics Relativistic global Code \citep{Baraka2021, Ben-Jaffel2021} where the transport of  plasma originating from the Earth has been included. This paper follows the work started in Barakat et al. ``The Sun-Earth-Moon Connection: I--3D Global Kinetic simulation'', referred to here as Paper I. Paper I concentrated on the simulation of the Earth's environment and on the modifications/validation of the model. The present paper concentrates on the Lunar environment. In the following, Section 2 quickly summarizes the parameters used in the simulation. Section 3 describes the Lunar plasma environment and the ion transport from the Earth to the Moon as simulated by the model. Section 4 discusses the results in light of previous observations, before the conclusion.
\clearpage
\justify 	 \section{Initial conditions and Simulation Model: IAPIC}\label{code}
The initial conditions are similar to the one in Paper I. More specific details on the formulation can be found in \citep{Baraka2021}. In order to observe the variations in plasma parameters, three distinct regions were selected within the simulation box to investigate how these parameters evolve. The values presented in Table \ref{moonparam} are expressed in code units, and the conversion from real-world values to code units is thoroughly documented in \citep{Baraka2021}
%\centering
\begin{table}[!ht]\caption{The plasma parameters were computed independently for two regions: the undisturbed solar wind at -24 to -18 \(R_E\) and the dayside magnetosphere, and the regions \(\pm 4 R_E\) on either side of the moon were considered to avoid positions where solar wind particles were eliminated. All quantities were averaged over the total number of cells considered to account for charge separation and species accumulation at Y=\(1R_E\) and Z=\(1R_E\).}\vspace*{0.1in} \label{moonparam}

\resizebox{\textwidth}{!}{\begin{tabular}{|l|s|a|q|}
	\hline
\rowcolor{yellow} parameters(averaged over \(4R_E)\)&Undisturbed SW	&  Lunar Dayside(\(-4R_E\))&Lunar Nightside(\(+4R_E\))  \\
\hline
Alfven velocity-\(v_A\) &     0.22&    0.1&     0.14\\
sound speed-\(c_S\) &    0.05&    0.05&    0.06\\
magnetosonic mach number-\(M_{ms}\) &     0.92&     0.62&     0.5\\
sonic mach number-\(M_S\)&3.84&      1.35&      1.33\\
 alfven mach number-\(M_A\)&     0.95&     0.70&     0.54\\
Plasma ion frequency-\(\omega_{pi}\)&    0.02&    0.02&    0.03\\
Plasma electron frequency-\(\omega_{pe}\)&     0.2&     0.26&     0.30\\
Plasma ion gyro frequency--\(\omega_{ci}\) &   0.01&   0.01&   0.01\\
Plasma electron gyro frequency--\(\omega_{ce}\)&     0.97&     0.65&     0.75\\
Ion Intertial Length-\(d_i\)&      22.21&      14.70&      19.15\\
Electron Intertial Length-\(d_e\)&      2.23&      1.92&      1.66\\
 gyro radius-\(r_i\)&     0.04&     0.23&     0.26\\
gyro radius-\(r_e\) &  1.e-3&   1.e-3&   1.e-3\\
Ion Debye Length-\(\lambda_{Di}\)&    0.02&    0.04&    0.07\\
Electron Debye Length-\(\lambda_{De}\) &   3.e-3&   3.e-3&   3.e-3\\
 Ion beta -\(\beta_i\)&  0.07&     3.41&      2.79\\
 Electron beta--\(\beta_{e}\)&     0.23&   0.6&     0.83\\
ompi/omci-\(\omega_{pi}/\omega_{ci}\)&      2.32&      5.19&      3.46\\
ompe/omce--\(\omega_{pe}/\omega_{ce}\)&     0.23&     0.40&     0.40\\
\(v_{thi}\)/\(v_{sw}\)&   3.5e-3&    1.3e-3&    1.8e-3\\
\(v_{the}\)/\(v_{sw}\)&   6.2.e-3&    7.1e-3&    8.2e-3\\ \hline
\end{tabular}}
\end{table}
\clearpage
\section{Simulation Results}\label{results}  
\justifying
During the full moon phase, the solar wind impinging on the far magnetotail impacts the lunar surface in various ways, including its electric charging. The present study focuses on the accumulation of solar wind densities, which are considered absorbed by the lunar surface by typical models. Not following that hypothesis leads to the decoupling of ions and electrons, resulting in charge separation and therefore in electric currents.\par
More precisely, the dayside of the lunar surface becomes positively charged, while the nightside becomes negatively charged, as illustrated in Figure (\ref{denzoom}). The UV radiation would increase the ionization rate on the dayside but are not taken into account in the present study and will deserve further investigation. \par

As depicted by the green arrows in Figure \ref{denzoom}, anti-corrections of solar wind densities occur in three distinct regions: the foreshock, in the dayside magnetosphere, and on both the lit and dark sides of the moon. Figure \ref{denzoom} (b) and (c) panels provide a close-up look at the ions and electrons drifting towards the day and night sides of the moon (centered for clarity at zero), respectively. \par Figure \ref{denboundary} shows the density when the moon is not positioned at 60\(R_E\) along the Sun-Earth line. The figure depicts ion densities (a) and electron densities (b) of the solar wind at \(\pm 2R_L\) in both dawn and dusk directions. The asymmetries between dawn and dusk regions in the deep magnetotail can be highlighted by comparing the two figures. If the moon is absent, the average densities at both dawn and dusk locations should be similar to those at the lunar location. It is important to mention that the total count of solar wind components is determined at the lunar surface within a spherical shell with a thickness of $0.1 R_M$, approximately 174 km, on both sides of the lunar surface. \par
The results of the simulation indicate that the number of ions on the dayside of the moon outweighs that of electrons, whereas the opposite is true on the night side of the moon. Table \ref{charges} presents an estimate of the overall count of ions and electrons, providing a comparison of the respective charges on each side of the moon. Previous research by \citep{Deca2015} suggested that the Lunar Magnetic Anomalies, mini-magnetospheres, could exacerbate these charge separations. However, this topic is beyond the scope of the current study. The phenomenon of lunar surface charging has been widely documented and discussed within the scientific community.\citep{Stubbs2007,Fatemi2014,Poppe2014,Halekas2014,Lue2018,Poppe2021a}. \par 

  Due to insufficient time for plasma to refill the wake with thermal velocities, typical lunar wakes often develop in the supersonic background plasma of the solar wind or remote magnetosheath. THEMIS observations have revealed that the field-aligned plasma flow within the lunar wake is highly organized when it is subsonic, particularly in cases where the plasma beta is low \citep{Xu2019}. The results indicate a preference for electrons to enter the wake from the dawn side of the Moon, while ions tend to refill the wake from the dusk side.  The computed charging percentages shown in Table \ref{charges} are consistent with the charging percentage reported in \citep{Poppe2021a}.\par

The densities (Figure \ref{wakeden}) and temperatures (Figure \ref{waketemp}) of ions and electrons at various distances from the lunar surface allow to obtain a comprehensive understanding of the plasma parameters within the lunar wake. Specifically, these values were extracted at the lunar surface, the terminator, and at distances of 3, 5, 7, and 9 \(R_L\) inside the wake. The wake had a width of \(\pm6 R_L\), extending equally on both sides of the dawn-dusk directions.\par
Figure \ref{wakeden}-a highlights the tracking of ion densities along the OX direction, starting from the lunar surface and extending up to 9 \(R_L\). The assumption of the plasma accumulating on the lunar surface causes the front flow of the plasma to shift towards a less resistive direction in both the dawn and dusk directions. Specifically, at the subsolar lunar surface, the plasma flow is halted. At the terminator, the ion densities are significantly higher in the dawn direction compared to the dusk direction.\par
 Inside the lunar wake, at a distance of  3\(R_L\), that plasma is slipping into the wake more prominently in the dusk direction, with a ratio of \(3:2\) compared to the dawn side. The average density of this slipping plasma is approximately   100 ions cm$^{-3}$.  Moving closer to the lunar surface, at 2\(R_L\)  there is an increase in ion filling, which is drifting towards the dusk direction. The rough density at this point is around 20 ion  cm$^{-3}$.  However, as we reach 7 \(R_L\), the density begins to fade away, and the number of ions reduces to approximately 6 ion  cm$^{-3}$. This value remains constant until 9  \(R_L\).\par
In Figure \ref{wakeden}-b, an interesting observation is made regarding the number of electrons at the lunar surface's subsolar point:  their count is close to that of ions. However, away from the terminator, a significant change in the ratio occurs, reaching an order of magnitude difference, with electrons outnumbering ions by a factor of 10.\par Furthermore, it is noticed that electrons tend to slip inside the lunar wake, showing a preference for the dawn side boundaries of the wake. The average number of electrons decreases from 300 to 5 electrons cm$^{-3}$  from 3\(R_L\) to 9 \(R_L\) within the lunar wake. The obtained results align with the analytical and hybrid-kinetic model simulations, as well as observations from the ARTEMIS mission, as reported in \citet{Gharaee2015}\par 
In Figure \ref{waketemp} (a) and (b), the temperatures of solar wind ions and electrons inside the lunar wake are depicted. A significant heating effect is observed, beginning at a distance of 7 \(R_L\) and continuing up to 9\(R_L\) The temperature remains elevated within this range.\par
 Near the lunar surface,  in Figure \ref{magdenwake}, the magnetic field components (\(B_x, B_y, B_z, and B_{tot}\)) along the ion densities are analyzed at two specific locations: the terminator (depicted in blue) and 3\(R_L\) inside the lunar wake (depicted in red). Ion densities are also shown at the lunar surface (in green).\par
A minor discrepancy in the magnetic field components can be seen in Figure \ref{magdenwake}-(b), with the exception of \(B_y\). This could be attributed to the significant disparity in the electric field at a distance of 3 times the lunar radius (3\(R_L\), since the \(B_y\) component at this distance exhibits symmetry both at dawn and dusk. It is noteworthy that this component exhibits a change in polarity inside the lunar wake, indicating a distinct magnetic field behavior within that region. The Magnetic field values at 3\(R_L\) [\(B_x, B_y, B_z, and B_{tot}\)] reads [2,-0.2, 5,5]nT. Looking at the solar wind ion density in Figure \ref{magdenwake}-(e), we observe an average density of \(\approx\) 10 ions cm$^{-3}$. However, comparing the density at the lunar surface to the density at the terminator, there is a noticeable flare-out of the plasma towards the anti-sunward direction.\par 
The magnetic field (a-d), electric fields (e-f), and solar wind velocity (i-m) on the moon's surfaces along three parallel planes: OX (\(\pm 5R_E\)), and at 2\(R_L\)  in both the dawn and dusk directions are also of interest to understand the effect of the Moon. Regarding the X-components in Figure \ref{emfvel} for magnetic field (\textbf{B}), electric field (\textbf{E}), and velocity (\textbf{V}), \textbf{B} is enhanced by a factor of 2 on the dayside, indicating bipolarity of \textbf{E}(between -3\(R_E\) and -2\(R_E\))  followed by a polarity change from 0.2 to -0.2 at -1.5 \(R_E\),  while \textbf{V} decreases until it reaches a stagnant state at the lunar surface. All dawn-dusk components exhibit asymmetry, therefore showing a model prediction that could be tested.\par
In terms of the Y-components, B shows enhancement, E remains nearly constant, and duskward velocity decreases until stagnation. The asymmetry of solar wind parameters is more pronounced on the dayside compared to the nightside.\par
As for the Z-components, \textbf{B} increases and exhibits almost symmetric behavior on the dayside, while E shows asymmetry and polarity reversal between --4\(R_E\) and -1\(R_E\), and V decreases from south to north until it reaches a stagnant point at the moon's position.\par
Concerning the overall plasma parameters, B is enhanced on both sides of the lunar surface, while a decrease is observed on the dusk side and constancy along the dawn side within the lunar wake. There is a significant asymmetry between dusk and dawn for \textbf{E} and \textbf{V}, while plasma velocity increases inside the lunar wake from 0-150 km/s.\par
Furthermore, Figure \ref{xyfields} illustrates two contour plots depicting the total magnetic field and total velocity in the equatorial plane, revealing the magnetic field enhancement and the corresponding solar wind ion velocity. In Figure \ref{xyfields}-a, the magnetic field enhancement is displayed on both the dayside and nightside of the lunar surface, with an average magnetic value reaching a maximum of 6nT (increased by a factor of 1.2 i.e. \citep{Wang2011}). On the other hand, Figure \ref{xyfields}-b shows velocity stagnation at the lunar surface ranging from 300 km/s to 0 km/s on the dayside, while inside the lunar wake in the equatorial plane, there is an increase of approximately 150 km/s.\par
In \citet{Kimura2008}, two-dimensional electromagnetic full particle code simulations were employed to successfully simulate the structure of the electric field near the lunar surface. Here, a three-dimensional electromagnetic relativistic global code \citep{Baraka2021,Ben-Jaffel2021} was used, by incorporating the formula in  \citep{Kimura2008} i.e., \(\mathrm{E_0=m_e v_e \omega_{pe}/q_{e}}\), where \(m_e\), is the electron mass, \(v_e\), is the electron velocity, \(\omega_{pe}\), the electron plasma frequency. and \(q_e\), is the electron charge,  to account for the background electric field.\par
As depicted in Figure \ref{denzoom}, and \ref{denboundary}, the phenomenon of charge separation gives rise to a potential difference, which we determined as an induced electric field \(E_0\). This electric field was utilized to standardize the solar wind electric field along both surfaces of the moon, specifically within the range of \(\pm 5R_E\)  Enlarging the moon's size by a factor of 5, equivalent to about 1.\(R_E\), does not impact the physical microscale of the process, which remains smaller than our grid size as shown in Figure \ref{evse0}.\par%
 {In Figure \ref{evse0}-A, the total Electric Field (E.F.) is plotted along three planes along OX, namely at \(Y=0\), \(Y=1RL\)(dawn) and at \(Y=-1RL\)(dusk) directions. The blue line represents the bulk E.F. The red line represents the background E.F. generated solely due to the charge separation. In order to account for an effective electric field, it is important to include the induced electric field (\(E_0\)) as a crucial component in our kinetic modeling.  It results in a bulk E.F. of 2.6\(E_0\) at \(Y=0\), 2.1\(E_0\) at \(Y=+1R_L\), and 2.6\(E_0\) at \(Y=-1R_L\), particularly at the terminator (\(Y=0\)). These results are consistent with those reported in \citet{Kimura2008} and the references therein.\par
{The electric field (Figure \ref{evse0}-B) in the present study has been normalized to the induced E.F. that results from the charge separation of both surfaces of the moon. Additionally, correlation can be performed between the total E.F. and the induced electric field: it reveals anti-correlations at \(Y=0\) with a correlation coefficient of C.C.=-0.06. At \(Y=+1R_L\), there is a positive correlation coefficient of C.C.=0.81. In contrast, at \(Y=-1R_L\), there is a weaker positive correlation with a correlation coefficient of C.C.=0.22.}\par
These findings have a remarkable consistency with \citet{Kimura2008}, indicating a complete agreement when there is a drop of the potential inside the lunar wake at around 2\(\mathrm{R_E}\) as in Figure \ref{evse0}b. Furthermore, the potential drop caused by surface charging is more pronounced in terms of  the potential drop associated with the lunar wake.\par
  Finally, the utilization of 3D simulation allows to obtain the values of  \(E_x, E_y, E_z, and E_{tot}\) enabling a comprehensive three-dimensional analysis of the magnetic field enhancement on both sides of the lunar surface. This enhancement is clearly illustrated in the left and middle panels of Figure \ref{emfvel}.\par
\begin{table}[!ht]\caption{	An estimation is made of the total number of positive and negative charges confined within a spherical shell with a thickness of 0.1$R_m$, approximately 174 km above the lunar surface. The density per unit volume is then calculated for both sides of the lunar surface. }\label{charges}
\begin{tabular}{|lccc|}			\hline
Charge sign	& Lunar dayside  & Lunar nightside& $\pm$ charge ratio \\ 
	\hline 		\hline \vspace*{0.1cm}
			Total \# of positive charges(ions)&\textbf{4.83$\times$10$^4$}&\textbf{2.74$\times$10$^4$}&\textbf{171\%}\\
			\vspace*{0.1cm} Total \# of negative charges(electrons)&\textbf{1.99$\times$10$^4$}&\textbf{7.04$\times$10$^4$}&\textbf{39.4\%}\\ 
			\vspace*{0.1cm} Relative charge&\textbf{2.84$\times$10$^4$}&\textbf{-4.30$\times$10$^4$}&\textbf{66\%}\\ 
			\multicolumn{4}{c}{\texttt{\textbf{\underline{Number density per unit volume on lunar surface}}}}\\[10pt]
		Number density	& Lunar dayside  & Lunar nightside& average density \\ 
		\textbf{$\rho_{i}$=$ N_{i} $/V}&\textbf{58$cc^{-1}$}&\textbf{33$cc^{-1}$}&$\overline{\rho_{i}}$=\textbf{45.5}$cc^{-1}$\\
		$\rho_{e}$=$ N_{e} $/V&\textbf{34$cc^{-1}$}&\textbf{85}$cc^{-1}$&$\overline{\rho_{e}}= \textbf{59.5}cc^{-1}$\\ \hline
		\end{tabular}
	\end{table}
{These simulations reveal a novel point regarding backstreaming ions at $\pm4R_E$ on both the lit and dark sides of the Moon. The present characterization of the backstreaming ions is consistent with previous studies \citep{Bonifazi1981, Baraka2021}. However, here, the backstreaming ions are not primarily reflected ions but instead diffuse ions with a small percentage of intermediate ions. To prevent any confusion, the reflection presented here is from the nightside of the Moon to its dayside; this is not a reflection from the surface of the Moon which has be demonstrated to be negligible \citep{Holmstrom2012}.
{Figure (\ref{anisotemp} -a) presents the co-plotted ion perpendicular temperature ($Ti_{\bot}$) and ion parallel temperature ($Ti_\parallel$) to study the temperature anisotropy in the near Moon during the full Moon phase, in the range of \(\pm 5 R_E\). Similarly, Figure \ref{anisotemp} -b presents the electron perpendicular temperature ($Te_{\bot}$) and electron parallel temperature ($Te_\parallel$) to study the temperature anisotropy. Previous studies, such as \citep{Gary2006,Chandran2011,Treumann2013,Karimabadi2014,Gingell2015}, have also reported temperature anisotropy in the Moon's vicinity. We find that $Ti_{\bot}$/$Ti_\parallel$ and $Te_{\bot}$/$Te_\parallel$ equal 9.2 and 7, respectively. The electron correlation coefficient is 0.9, and the ion correlation coefficient is 0.75. It has been reported in studies such as \citet{Samsonov2012,Grygorov2017} that the backstreaming ions (Figure \ref{backstreaming}) can affect the temperature anisotropy configuration.}\par 
{Table \ref{backvelocity} presents the bulk speed of the backstreaming ions, along with their corresponding thermal velocities and the solar wind kinetic inflow. Notably, the thermal velocity of the backscattered ions on both sides of the Moon is higher than the bulk flow and the solar wind inflow speed. Additionally, the speeds are slower at the day side of the Moon than at night.}\par 
\begin{table}[!ht]\caption{The values of the solar wind ions for inflow, thermal, and bulk backstreaming speeds are listed for both the dayside lunar surface and inside the lunar wake.  }\label{backvelocity}
\begin{tabular}{|l|c|c|c|}
			\hline
Speed& $V_{SW} $-Lunar dayside  & $V_{SW} $-Lunar nightside & Night/day $V_{SW} $ ratio\\ 
			\hline 		\hline \vspace*{0.1cm}
$V_{backstreaming_{(bulk)}}$
&\textbf{58.6$ ~~km.sec^{-1}$}&\textbf{136.4$ ~~km.sec^{-1}$}&\textbf{2.3}\\ \hline 	
\vspace*{0.1cm} $V_{backstreaming_{(thermal)}}$&\textbf{99$ ~~km.sec^{-1}$}&\textbf{186$ ~~km.sec^{-1}$}&\textbf{1.9}\\ \hline 	
\vspace*{0.1cm} $V_{SW_{(inflow)}}$&\textbf{80.6$ ~~km.sec^{-1}$}&\textbf{110$ ~~km.sec^{-1}$}&\textbf{1.4}\\ \hline
	\end{tabular}
\end{table}
\justifying
\paragraph*{In Summary:}
The charge separation on the lunar surface gives rise to an induced electric field, which is closely associated with ions backstreaming. This backstreaming phenomenon, in turn, is linked to the temperature anisotropy observed in the lunar environment. Figure \ref{chart} depicts the findings related to the connection between the Sun and Earth, the coupling between the magnetosphere and ionosphere, as well as the interaction between the solar wind and the Moon. This sketch offers a visual depiction of the findings from the study and illustrates their interconnections.\par
\clearpage
\section{Discussion on observational implications}\label{discuss}
\justifying
{This discussion focuses on the Lunar surface charging and on the plasma parameters within the Lunar wake when the Moon is positioned solely along the Sun-Earth line, i.e., during a Lunar eclipse and inside the extended magnetotail of the Earth}.\par 
\subsection{Lunar Surface Charging}
{ Contrarily to previous studies such as \citet{Poppe2014}, the present self-consistent simulation did not remove or absorb solar wind particles upon collision with the lunar surface. These particles are instead allowed to accumulate  and build up on the lunar unmagnetized barrier. By doing so, we observed charge separation(see Figure \ref{denzoom}), which led to a dominance of positively charged particles on the dayside of the Moon and a prevalence of dominance of negatively charged particles on the nightside. One should be careful with this approach, because measurements show that protons are absorbed by the lunar surface \citep{Holmstrom2012}, which is what partially leads to the creation of water with the regolith, future work will improve on this interaction. }\par
 { This approach aimed to account for the total number of charged particles on both Lunar surfaces, as demonstrated in Table \ref{charges}. Typically, the Debye length determines the distance at which charges are separated or screened. In our case, the charges of piled-up solar wind plasma considered at step 3700 $\Delta t$ were shielded from the influence of distant particles closer to them. As a result, the total Debye length in the piled-up plasma is given by  \begin{equation}\label{debye}
	 \lambda_{D_l}=\sqrt{\dfrac{K_B}{4\pi }\Sigma_l \dfrac{ T_l}{q_l^2 n_l}}
\end{equation}
as reported in \citet[Eq. 120]{Verscharen2019}. }\par 
{The charging of the Lunar surface in this complex and non-uniform environment can be attributed to various complex current systems, including photoemission of electrons, plasma electrons, plasma ions, and secondary electrons resulting from surface ionization. \citep[and references therein][]{Halekas2005,Stubbs2007,Collier2014,Verscharen2019,Chandran2022}. The amount of charging depends on several factors, including the density of the species present (\(n_0\)), as well as the temperatures of the ions and electrons (\(T_i, T_e\), respectively), and the bulk flow velocity of the solar wind (\(v_{i,e}\)), as discussed in \citet{Stubbs2014}.}\par 
{The present simulation results, obtained using the PIC code, can be compared with other simulations and observations conducted at local and global scales (\citep{Halekas2011}; \citep{Poppe2014}; \citep{Deca2015}). Since the present study aims to explore the charging of the lunar surface by analyzing the accumulation of solar wind particles on both sides of the Moon, the density at which the correlation between ions and electrons breaks up can be used to estimate the total number of charges on both lunar surfaces Figures \ref{denzoom} and \ref{denboundary}. This breaking of correlations between ions and electrons densities indeed leads to charge separations due to topological differences(\citep{Bale1997}. The total number of positive and negative charges in a spherical shell with a thickness equivalent to 0.1 \(R_L\) (\(\approx \)175 km) above the lunar surface has been computed (Table \ref{charges})}. { This suggests that the dayside of the Moon is positively charged, in agreement with \citet{Kimura2008}, while the nightside of the Moon is negatively charged \citep{Stubbs2014}. These results are also comparable with measurements \citep{Halekas2002}. It is worth noting that the linear density at both the dawn and dusk positions in Figure \ref{denboundary} provides an estimate of the density measured along the OX direction of the Moon when it is not in its current position.}\par
{Despite previous studies and references indicating a predominance of negative charge in the lunar nightside which could implies that no positive ions are transported to it, the SELENE mission \citep{Nishino2010} demonstrated the entry of ions in the lunar wake. The present simulation aligns with these observation:  Table \ref{backstreaming} shows that approximately 39\% of ions detected within the lunar wake are solar wind particle. This result highlights the versatile and inclusive nature of the kinetic simulation of this complex interaction.}
{The potential differences between the two surfaces of the Moon cause a charge separation that leads to the creation of an induced electric field. To determine the background electric field, we utilized the formula outlined in \citet{Kimura2008}, which is given by \(E_0=m_e v_e \omega_{pe}/q_e\).}\par { As depicted in Figure \ref{evse0}, the electric field intensity along the subsolar Moon is normalized to the background electric field and found to be \(E=2.6E_0 \)
. Similarly, at dawn and dusk (terminator) directions, the values are \(2.1E_0\) and \(2.6E_0\), respectively. These results align with those reported in \citet{Kimura2008}, which indicate a measurement of \(2.2E_0\) at the terminator. }\par
This result should be put in perspective with other magnetospheric environments: in planets such as Jupiter, the magnetotail current system is influenced by the planetary magnetic field. However, in the cases of Venus and Mars, the magnetotail current system is primarily formed by the solar wind's magnetic field. The Earth's magnetotail, on the other hand, can be considered an intermediate case, where the specific configuration depends on the geomagnetic conditions and the dynamics of the magnetotail itself. It is a region connected to the Earth and is populated by hot, rarefied plasma. This is a complex intermediate case study (i.e. \citet{}{Xu2018}). % than can be uniquely studied by our code.
Therefore, to accurately account for the effective electric field near the lunar surface, it is essential to employ kinetic modeling, which considers the backstreaming ions and effectively captures the induced electric fields resulting from charge separation. The present results are therefore one of the rare examples of the whole consistent simulation of the system.}\par
\clearpage
%%%%%%%%%%%%%%% section 2 in discussion
\subsection{Plasma Parameters Within the Lunar Wake}

{Figure \ref{magdenwake} presents the magnetic field components, total magnetic field, and density of solar wind ions observed at the Lunar terminator and \(3R_E\) inside the Lunar wake. This figure is evaluated against \citet[Fig. 4]{Poppe2014}, where the  two events captured by ARTEMIS are compared  with a model. The present simulation results are shown for a single time step (3700\(\Delta t \approx 40 min\) ), whereas, in Figure 4 of \citet{Poppe2014}, the data was captured over 50 minutes.}
{The magnetic field components inside the lunar wake at coordinates [63,0,0]\(\mathrm{R_E}\) are computed to be [bx,by,bz]=[2.30,-0.07,4.80]nT. A comparison with  Table 1 of \citet{Poppe2014} is unfortunately limited to one specific event due to the size of the present simulation box. The March 2011 events has the spacecraft's location within the simulated region. The model at coordinates [56.9, 19.8, 3.1]\(\mathrm{R_E}\), gives a magnetic field of [-2.86,-0.87,2.17]nT to be compared with the measured value of [-1.5, 0, -1]nT \citep{Poppe2014}. The modeled values are almost double the observation; the reason for this discrepancy is related to the different solar wind speed, and initial IMF values. A more detailed study of this exact observation would involve running the model with the exact parameters of that day and is out of the scope of the present study.
The Lunar Prospector data show an increase of the magnetic field at the center of the wake followed by a decrease near its boundaries\citep{Akimov2012,Poppe2014}. These results do not compare directly to our simulations since it takes only a phase of the Moon and do not consider the present alignment. However, Figure \ref{emfvel} shows such an increased magnetic field near the center of the wake, at a total value of 5.5 nT near the center while it goes to 4 nT four Moon radius's away during the full moon phase.
Figure \ref{emfvel}-a,e,\&i presents an increase in the magnetic field on the dayside Lunar surface (depicted in red) along the X-components of the plasma parameters. This increase corresponds to bipolarity (normalized to induced \(E_0=m_e v_e \omega_{pi}/q_e\)) in the electric field, which is a direct consequence of charge separation and induced potential, as well as the stagnated solar wind velocity.\par
The remaining panels in Figure \ref{emfvel} demonstrate the variations in the magnetic and electric fields, as well as solar wind velocities, across the Y, Z, and total components.  The Moon is situated at a distance of
 \(60R_E\). Magnetic field measurements are presented in units of nT, while velocities are given in \(km/s\), The electric field values are normalized based on the induced background field that arises due to charge separations.\par

Higher solar wind ion velocity/temperature is observed at $\pm$5 R$_E$ as shown in Table~\ref{backvelocity}. This suggest that external forces, energetic inputs, or non-equilibrium conditions increase these velocities \citep[i.e.,][]{Futaana2012}. It is to be noted that the solar wind not only provides electrons and ions but also 20\% of energetic neutral atoms \citep{Vorburger2016} whose effects are out of the scope of the present study.
 In order to address the asymmetry between dawn(blue) and dusk(green), the correlation coefficients were computed for the magnetic and electric fields, as well as solar wind velocities, along the dawn-dusk directions. The analysis revealed that the correlation coefficients for  \(B_x\) and \(B_y\) were negative at both dawn and dusk, with values of -0.76 and -0.11, respectively. In contrast, the correlation coefficient for \(B_z\) C.C. is 0.43 was positive at both dawn and dusk, with a value of 0.43. Furthermore, the total magnetic field was found to be positively correlated at both dawn and dusk, with a coefficient of 0.44.\par
It is noteworthy that the significant disturbances observed in the magnetic field components near the surface of the Moon are of crucial importance since they are directly associated with the interaction between the lunar surface and the solar wind. This can aid in distinguishing the effects of various factors, such as photoelectron radiation and lunar magnetic anomalies, as reported in previous studies (\cite{Zhang2020,Deca2015}).\par
The correlation coefficients for the electric field components were calculated as follows: \(E_x\), \(E_z\), and \(E_{tot}\) exhibited negative correlation, with coefficients of -0.06, -0.78, and -0.12, respectively. However, the \(E_y\) component showed positive correlation, with a coefficient of 0.7. In addition, the velocities measured at both dawn and dusk were found to be positively correlated, with correlation coefficients of [\(v_x, v_y, v_z, v_{tot}\)] equal to 0.3, 0.82, 0.57, and 0.09, respectively.\par
Apart from highlighting the dawn-dusk asymmetry in the region surrounding the Moon, the dawn (green) and dusk (blue) magnetic and electric fields, as well as velocities, can also serve as a reference level for investigating plasma parameters in the magnetotail, assuming that the Moon is absent from its current location. Thus, these dawn-dusk asymmetric parameters can be employed to analyze the intricate current system in the magnetotail, up to a distance of 65 \(R_E\). \par
To better visualize the magnetic field enhancement on the lunar dayside and within the wake, we have presented 2D contours in equatorial planes taken at \(\pm 5R_E\) in both the day/night and dawn-dusk directions of the Moon. The Moon is centered at zero as shown in Figure \ref{xyfields}. In Figure \ref{xyfields}-a, it is evident that the magnetic field is amplified on the illuminated side of the Moon, and its value can be calculated using the formula \(\Delta B=(B_{moon}-B_{sw}/B_{sw} \))  i.e.\ \citep{Liuzzo2021}. This formula yields a value of (\(>\)1.) for the enhancement factor.\par
\justifying
{Figure \ref{evse0}-A shows the total Electric Field (E.F.) plotted along three planes, OX (at Y=0, Y=$+1R_L$(dawn), and at Y=$-1R_L$(dusk) directions. The bulk E.F. is shown in blue, while the background E.F. generated only due to charge separations is shown in red. The E.F. is considered at $\pm 5R_E$ at both the day and night sides of the lunar surface. It is demonstrated in Figure \ref{evse0} that the charge separation takes place at the lunar surface. The charge separation results in the induced additional E.F. ($E_0$), which is equal to \(E_0=m_e v_e \omega_{pe}/q_0\). To consider the effective E.F.,\(E_0\) should not be ignored, which can only be obtained by kinetic modeling. It is shown that the bulk E.F.=2.6\(E_0\) at Y=0, 2.1 \(E_0\) at \(Y=+1R_L\), and 2.6\(E_0\) at \(Y=-1R_L\) averaged over in the vicinity of the lunar surface, especially at the terminator \(Y=0\). These results are consistent with those reported in \citet[and references therein][]{Kimura2008} . As seen in Figure \ref{denzoom}, positive charges dominate the lunar dayside, and the nightside is dominated by negative charges, which also agrees with \citep{Kimura2008}. It is noteworthy that the Correlation Coefficients (C.C.) between these two field components are as follows: \(C.C.=0.71\) along \(Y=0\), 0.83 at \(+R_L\), and 0.54 at \(Y=-1R_L\), respectively}\par 
{Backstreaming ions are present at the lunar surface and within the Lunar wake, according to  \citet{Bamford2012}. This study found that the Electric Field (EF) deflects the incoming solar wind ions and that Chandrayaan-1 has observed backstreaming ions. In Figure \ref{evse0}, we present the effective electric field at both lunar surfaces, which may be responsible for deflecting the incoming solar wind.}\par {Additionally, it has been reported that the solar wind can be absorbed by the lunar surface, backscattered, or cause atoms to be removed from the Lunar regolith by sputtering or desorption \citep{Dandouras2023}. Some studies have reported that a large percentage of backscattered solar wind ions can result in the formation of Energetic Neutral Atoms (ENA), which have been observed by the Interstellar Boundary Explorer (IBEX) \citep{Allegrini2013}. Our Electromagnetic PIC Model does not account for ENA emissions, so the information presented in Figure \ref{backstreaming} only pertains to charged solar wind ions.}\par
{It should be highlighted that the backstreaming ions were not only observed in our simulations, but we were also able to analyze and categorize them based on the ratio between the thermal speed and bulk speed of the solar wind. Specifically, the backstreaming ions observed at both lunar surfaces were identified as being diffuse rather than reflective, which aligns with the criteria previously reported by \citet{Baraka2021}}\par
{Our simulation resulted in a solar wind bulk flow comprising roughly 23\% of backstreaming ions; the analysis accounted for the kinetic effects of these ions. Figure \ref{backstreaming} showcases the identification of backstreaming ions in the vicinity of the Moon, which are attributed to temperature anisotropy, as depicted in Figure \ref{anisotemp}.}
{In brief, the present simulation allowed the solar wind to accumulate on the lunar surface. This resulted in charge separations that generated a potential difference and induced an additional electric field component. In the absence of the lunar magnetosphere, the electric field deflected the incoming solar wind, leading to backstreaming ions at the lunar surface, which was associated with temperature anisotropy. This approach provides complementary information to potential future  microscale studies of the Moon that investigate the influence of Lunar surface roughness on volatile sources and sinks, abundance, and evolution in the Lunar environment (as discussed in \citet{Davidsson2021} and \citet{Grumpe2019}), as well as photoelectric emission \citep{Mishra2020}}.\par
\clearpage
\section{Conclusion and Future Work}\label{conclusion}
{In these two studies, we have conducted a comprehensive analysis of the solar wind interaction with the lunar surface, taking into account complex current systems originating from various sources. This research presents the first-ever kinetic simulation of the Sun-Earth-Moon system, focusing specifically on a single scenario where the Moon is situated within Earth's magnetotail.}\par
Our findings reveal that the dayside of the lunar surface is predominantly influenced by positively charged solar wind ions, while the night side experiences a higher concentration of negatively charged electrons, arising from charge separation. This relationship is clearly illustrated in the flow chart provided in Figure \ref{chart}.\par { Importantly, our results, more detailed in Paper I,  demonstrate that, under these conditions, Earth's magnetosphere does not protect the Moon from the solar wind, suggesting that the potential water creation process on the lunar surface, driven by the solar wind, is not hindered by Earth's magnetospheric presence.}\par
{Moreover, our investigation has enabled us to explore numerous plasma parameters within the lunar wake. These include plasma ion and electron densities (Figure \ref{denzoom}), plasma ion and electron temperatures (Figure \ref{waketemp}), the Interplanetary Magnetic Field (IMF) at the terminator and within the wake (Figure \ref{magdenwake}), backstreaming ion characteristics (Figure \ref{backstreaming}), solar wind temperature anisotropy (Figure \ref{anisotemp}), and the effective electric field accounting for the induced electric field resulting from charge separation (Figure \ref{evse0}). These critical aspects of kinetic simulations offer valuable insights into the environmental conditions surrounding the Moon.}\par
{Future work will involve placing the Moon in various locations, including those outside of Earth's magnetotail, examining the effects of different solar activity levels, and improving the modeling of Earth's polar escape to better understand the transport of oxygen ions to the Moon in these conditions. A more in-depth understanding of the lunar plasma environment will significantly benefit upcoming missions, particularly those enabled by the Artemis program and the Moon to Mars initiative \citep{Dandouras2023}.}
\clearpage
\singlespacing
\bibliography{sw-moon.bib}\label{ref}
\clearpage
\section{Acknowledgment}\label{ack}
This research was carried out at the National Institute of Aerospace, Science Sytems and Applications Inc., Langley Research Center, and partially at the Jet Propulsion Laboratory, California Institute of Technology, under a contract with the National Aeronautics and Space Administration Contract Number 80NM0018D0004. We express our gratitude for the valuable code contribution provided by Lotfi ben Jaffel. We gratefully acknowledge the assistance provided by Bjorn Davidsson, Larry Paxton, Richard Barkus, Iannis Dandouras, and Enrico Piazza for their immeasurable support and valuable discussions on the issues and progress of this research.
\clearpage
\section{Figures}\label{Figs}
%%%%Figure 1
\begin{figure}[ht]
	\centering
	\includegraphics[width=0.9\linewidth]{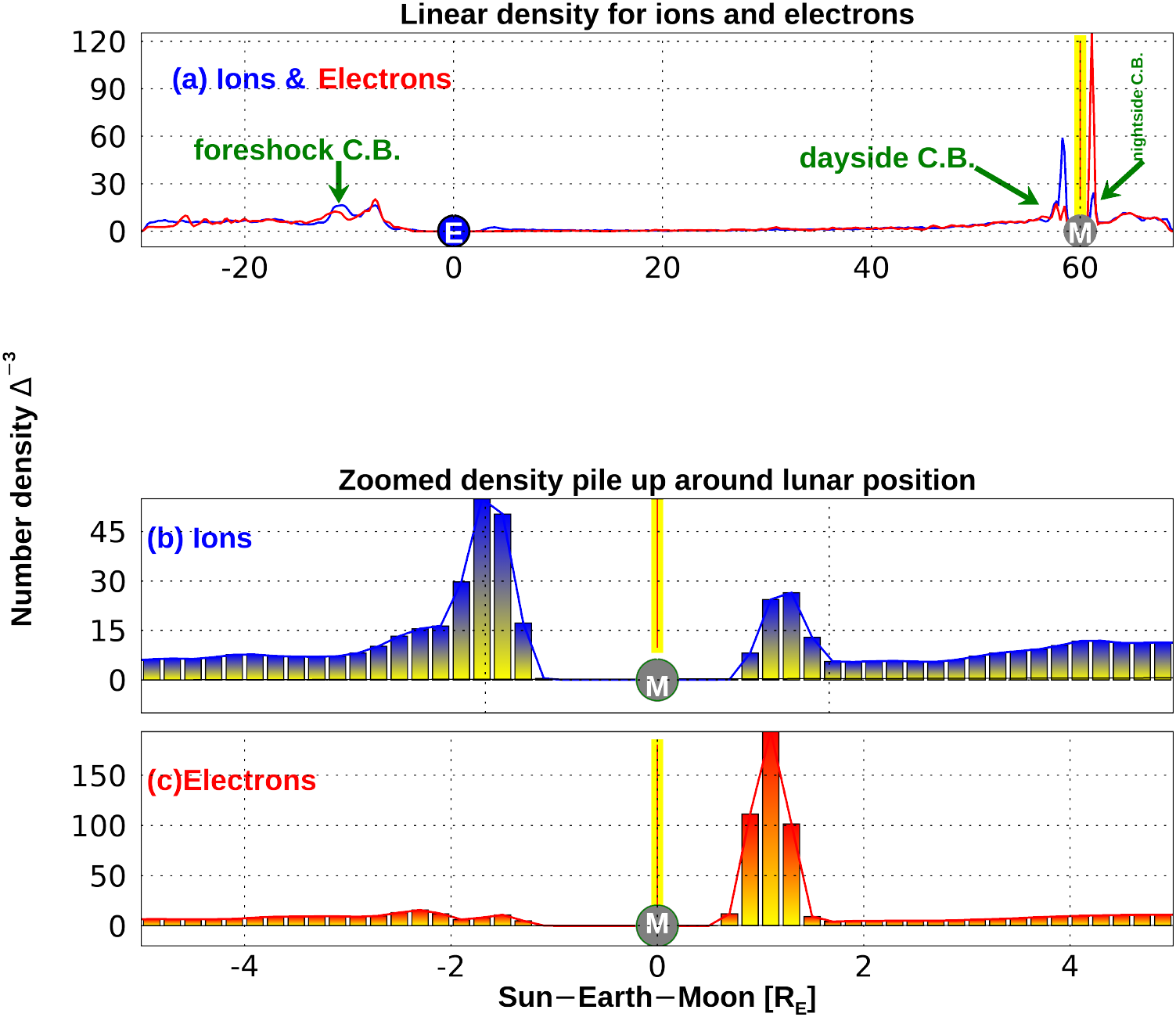}
	\caption{\fontfamily{cmtt} {The densities of ions and electrons in the axis of the Sun-Earth-Moon system are illustrated in the figure below. The areas where the potential is created are also indicated. The zoomed-in panels (b) and (c) show the ion and electron densities in the region around $\pm 5R_E$ from the Moon(centered at zero). The difference in charge accumulation between the Lunar dayside and nightside is highlighted in these panels.}}
	\label{denzoom}
\end{figure}
%%%%Figure 2
\begin{figure}[ht]
	\centering
	\includegraphics[width=0.9\linewidth]{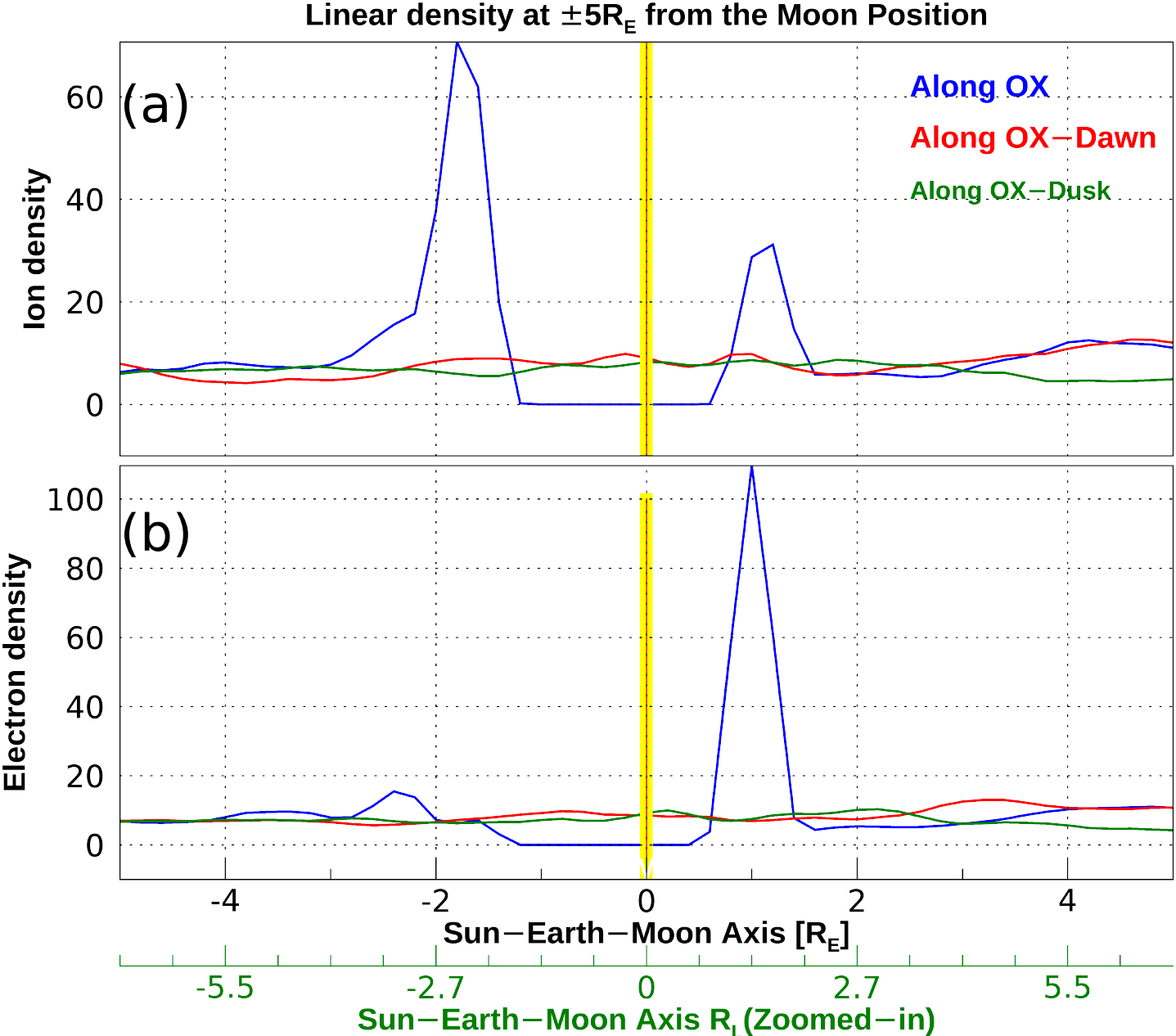}
	\caption{\fontfamily{cmtt}\selectfont {The distribution of SW (solar wind) particles is shown in the figure below, plotted along OX at $\pm 5R_E$ on the day and night sides of the Moon. The distribution is also shown along three planes: the plane of the planet's position and the dawn and dusk planes at $\pm 2R_M$ laterally. Panel (a) illustrates that ions outnumber electrons on the lunar dayside, while panel (b) shows that electrons outnumber ions on the night side. This phenomenon is caused by charge separation on both sides of the lunar surface. Specifically, the dayside lunar surface has a relative positive charge of $2.84 \times 10^4$, whereas the dark side of the Moon has a negative charge of $-4.3\times 10^4$ (see Table \ref{charges}).})}
	\label{denboundary}
\end{figure}
%%%%Figure 3
\begin{figure}
	\centering
	\begin{tabular}{cc}
		\includegraphics[width=0.49\linewidth]{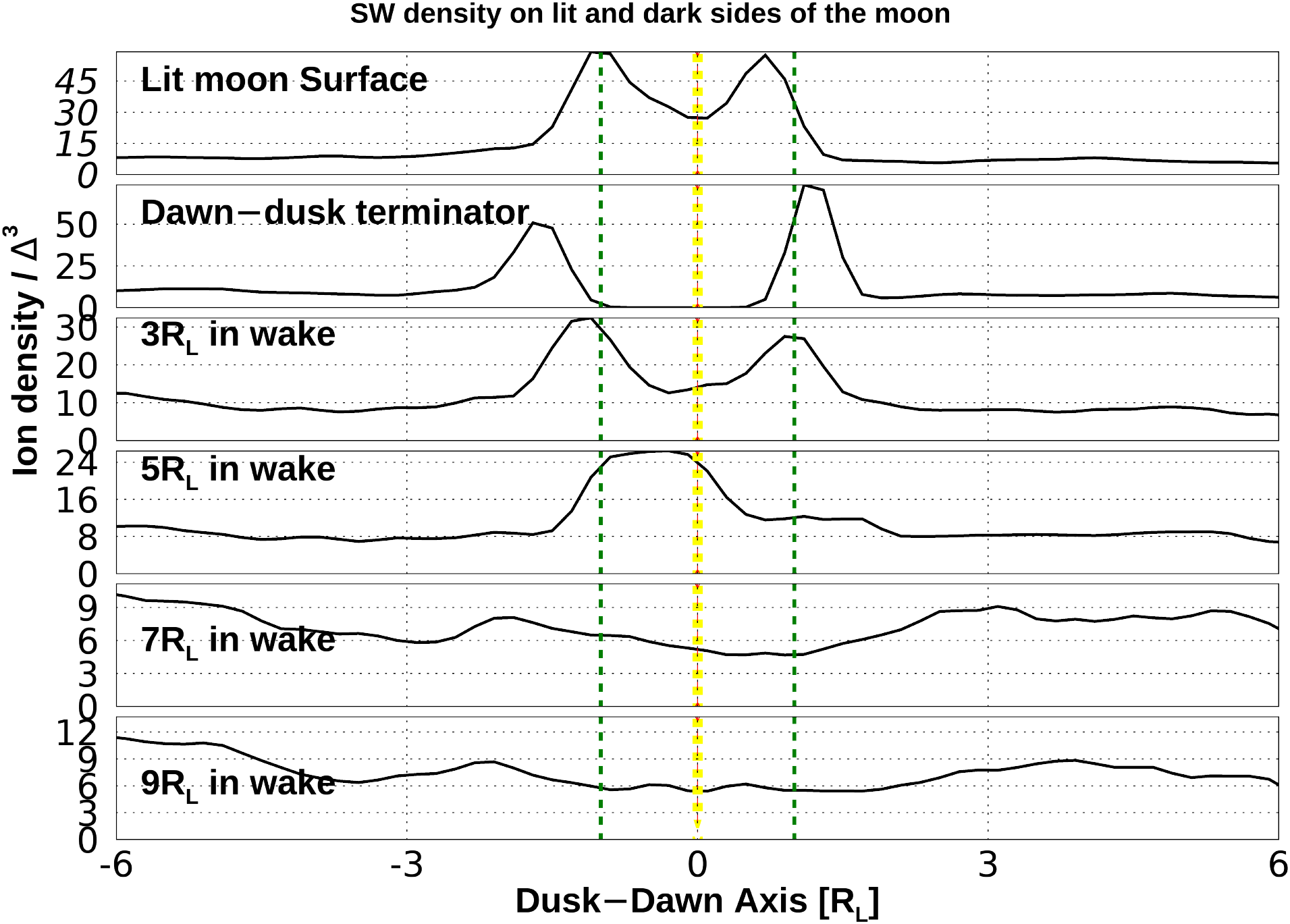}	&  	\includegraphics[width=0.49\linewidth]{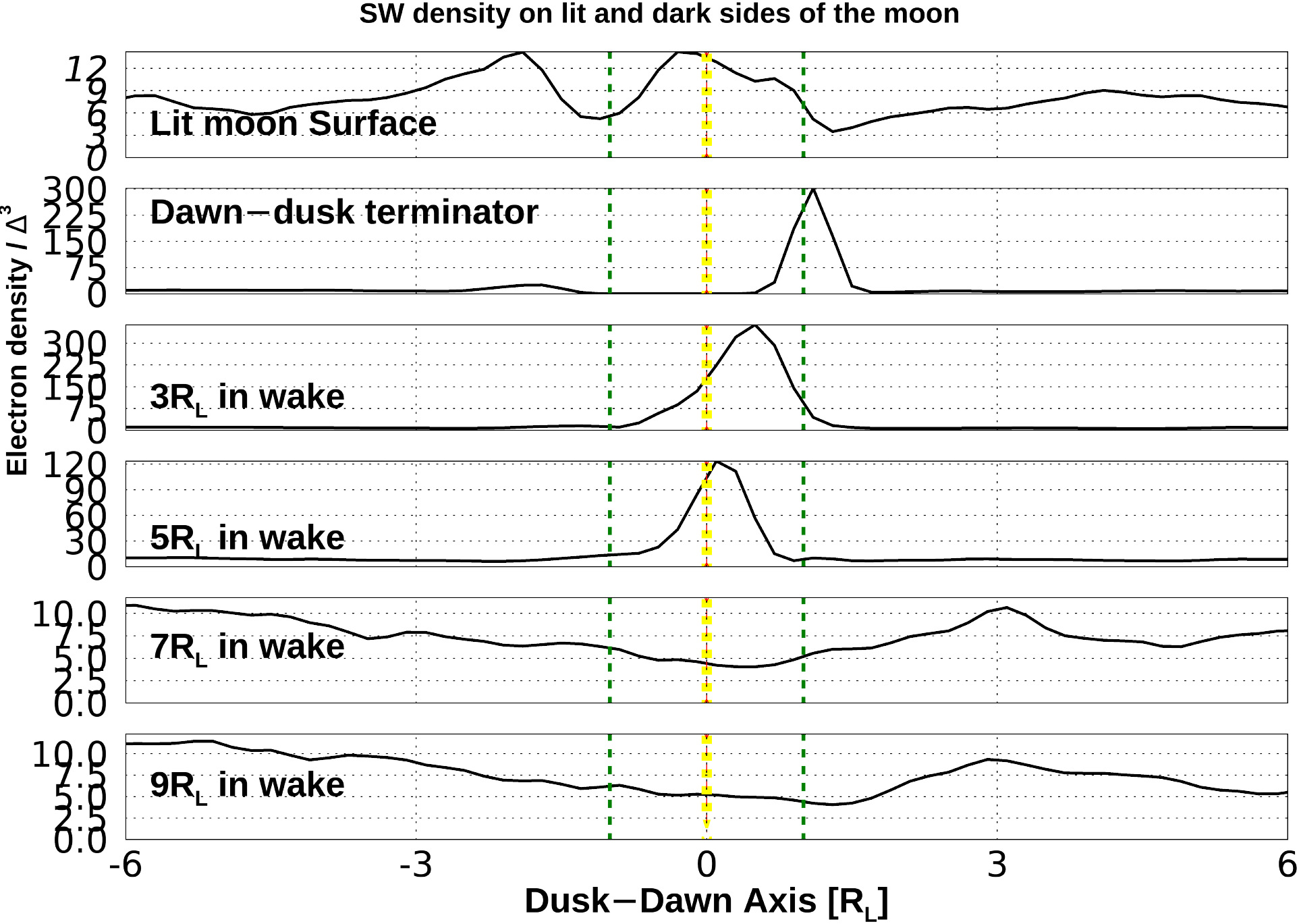}\\
		A&B\\
	\end{tabular}
	\caption{\fontfamily{cmtt}\selectfont {The figure  depicts a comparison between the profiles of ion (A) and electron (B) densities at the position of the Moon ($60R_E$) at various cuts along OX. These cuts are taken at the lit surface of the Moon, the dawn-dusk terminator, and 3, 5, 7, and $9R_M$ from the Moon, respectively. The densities are averaged over $1R_E$ in the OZ-direction and measured in the dusk-dawn direction at both sideways by $\pm6R_M$. The figure shows the refilling of the lunar wake by solar wind particles. Ions mostly enter from the dusk side, while electrons mainly enter from the dawn side. This phenomenon is directly caused by the charge separation at the lunar surface and the resulting potential difference.}}
	\label{wakeden}
\end{figure}
%% Figure 4
\begin{figure}
	\centering
	\begin{tabular}{cc}
		\includegraphics[width=0.49\linewidth]{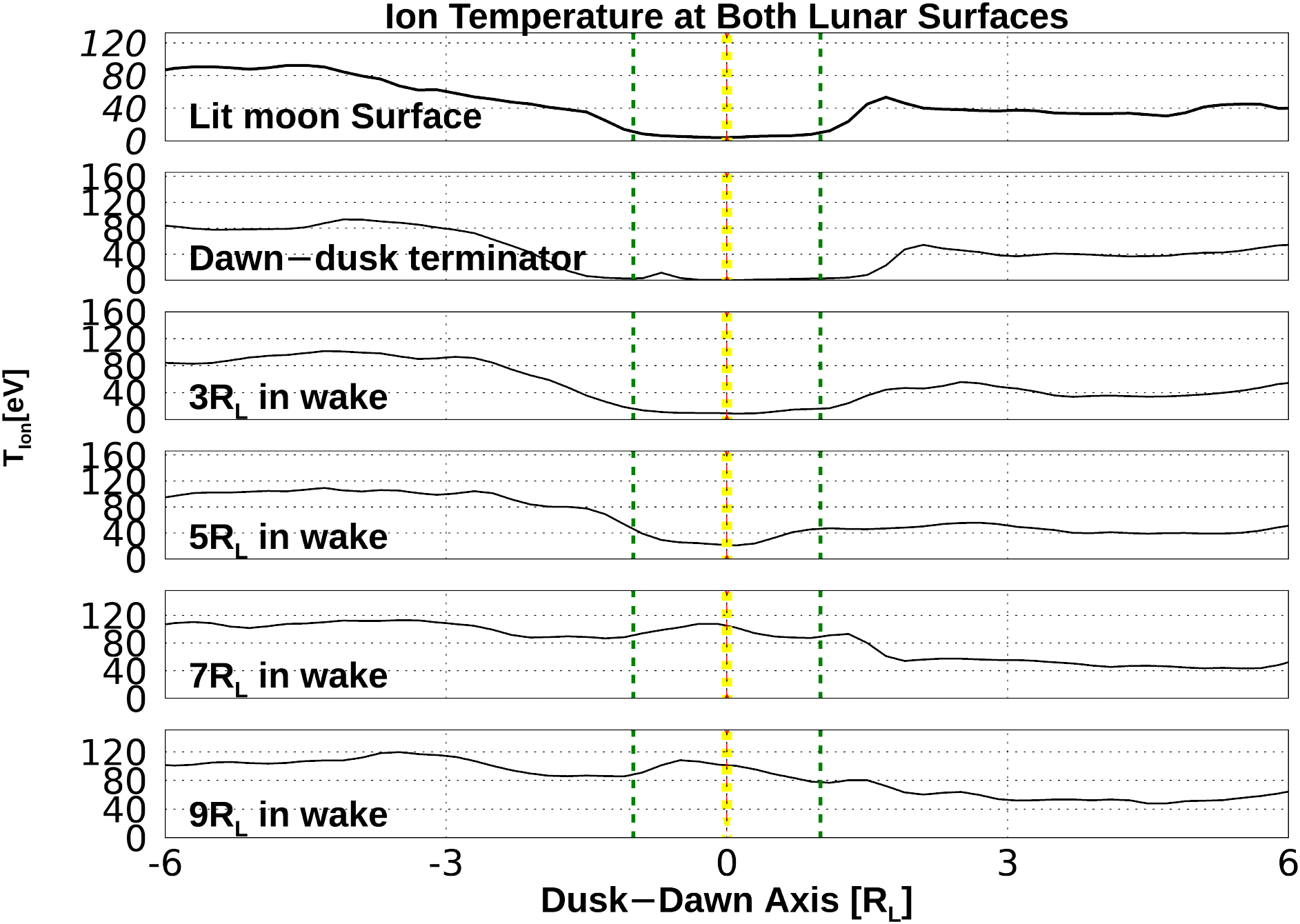}	&  	\includegraphics[width=0.49\linewidth]{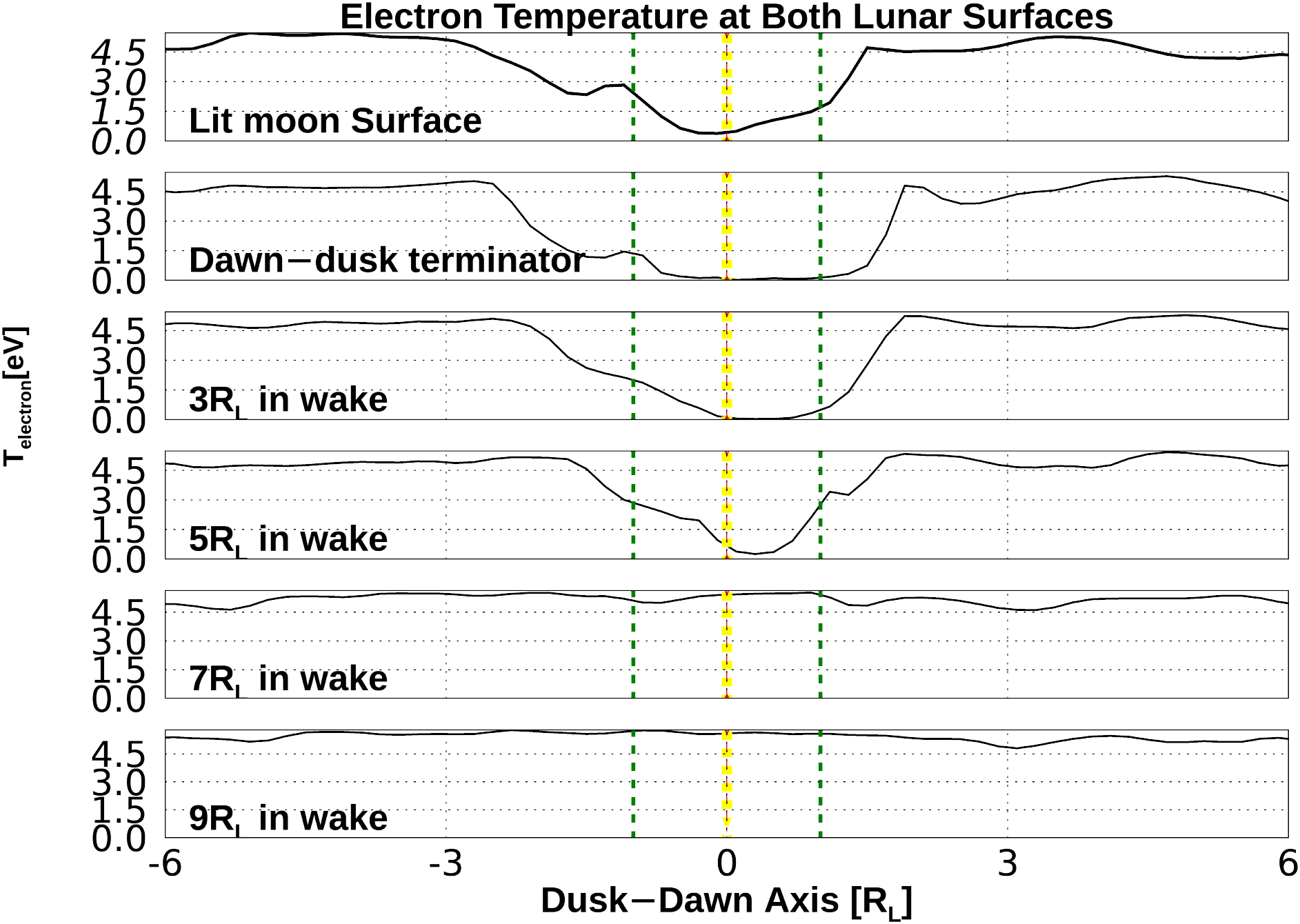}\\
		A&B\\
	\end{tabular}
	\caption{\fontfamily{cmtt}\selectfont {The figure  displays a comparison between the temperature profiles of ions (A) and electrons (B) at the position of the Moon ($60R_E$) at various cuts along OX. These cuts are taken at the lit surface of the Moon, the dawn-dusk terminator, and 3, 5, 7, and $9R_M$ from the Moon, respectively. The temperatures are averaged over $1R_E$ in the OZ-direction and measured in the dusk-dawn direction at both sideways by $\pm6R_M$. The figure depicts the refilling of the lunar wake by solar wind particles. Ions mainly enter from the dusk side, while electrons mainly enter from the dawn side. This phenomenon directly results from the charge separation at the lunar surface and the resulting potential difference. This figure corresponds to Fig. \ref{wakeden}.} }
	\label{waketemp}
\end{figure}
%%% Figure 5
\begin{figure}[ht]
	\centering
	\includegraphics[width=0.49\linewidth]{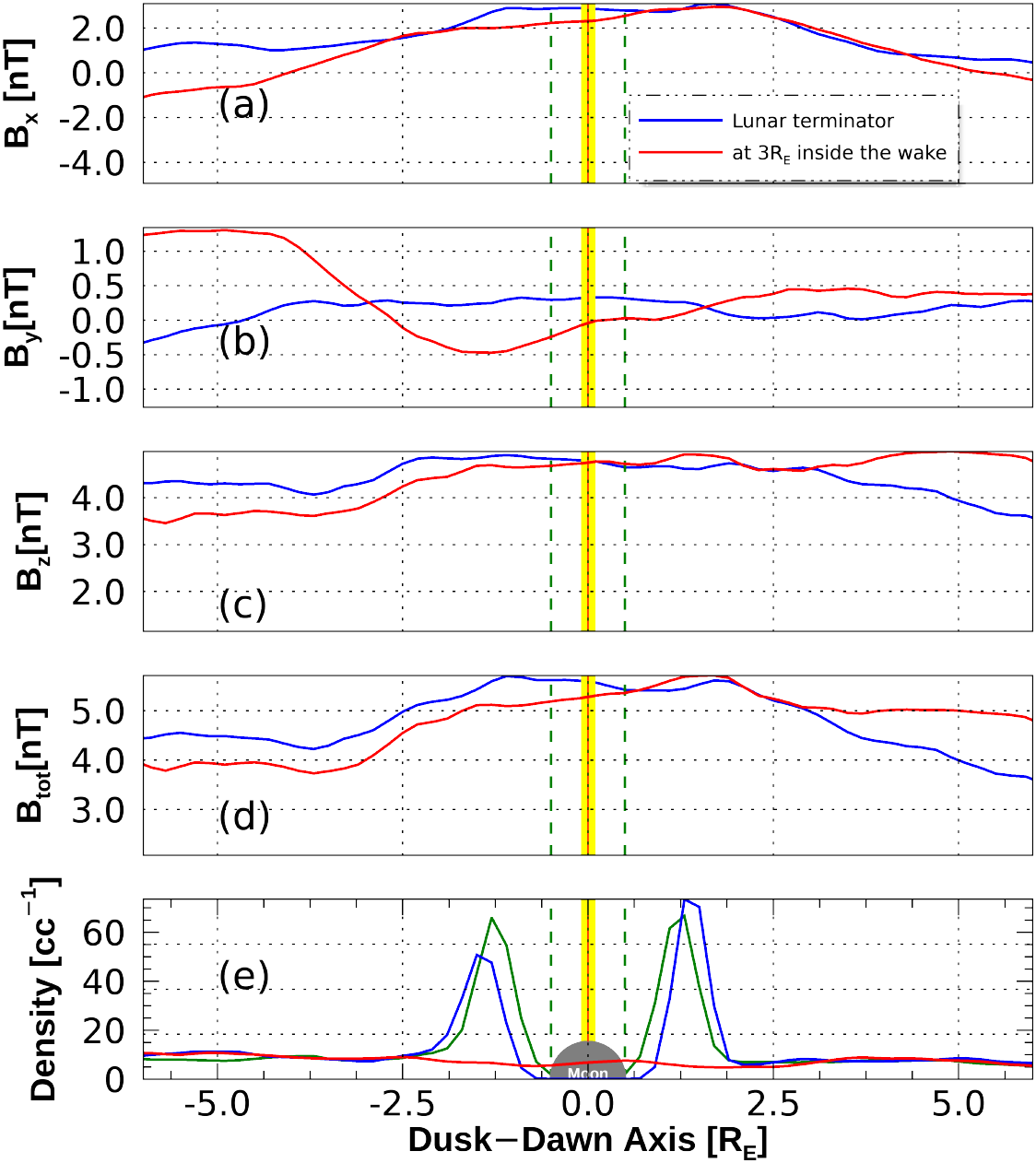}
	\caption{\fontfamily{cmtt}\selectfont {This Figure depicts the components of the magnetic field, the total magnetic field, and the density at the lunar terminator and within the Lunar wake. The boundaries of the Moon are marked by vertical green dashed lines. A difference in the density plots is noticeable between the linear graphs shown in green (taken at the lunar surface) and blue (taken at the terminator), revealing the flaring out of the density flow around the Moon. The findings are evaluated against the outcomes shown \citep[Fig. 4]{Poppe2014}.}}
	\label{magdenwake}
\end{figure}
%%%%Figure 6
\begin{figure}[ht]
	\centering
	\includegraphics[width=0.9\linewidth]{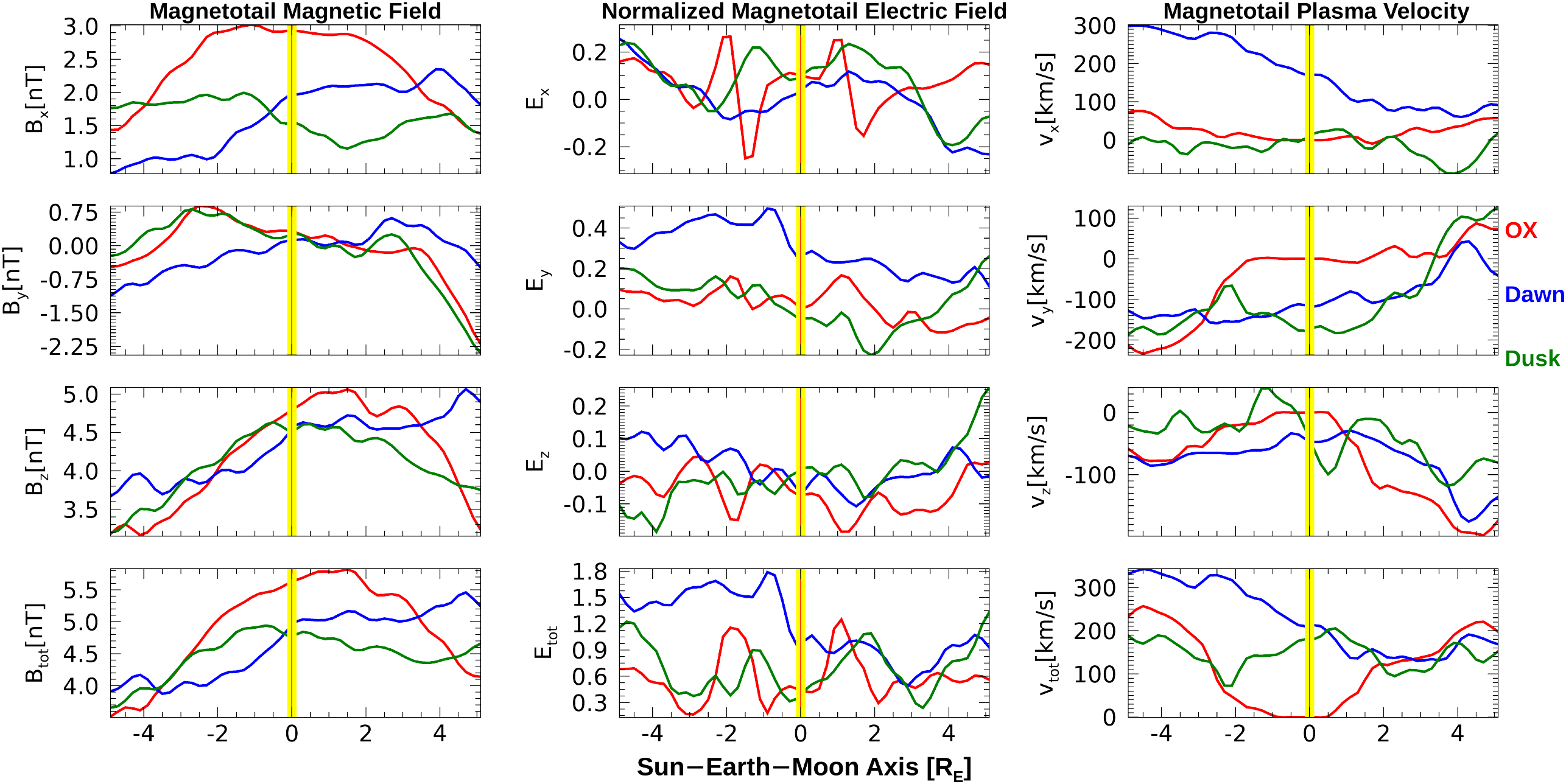}
	\caption{\fontfamily{cmtt} {This figure depicts the magnetic(a-d), electric field(e-h), and solar wind velocity(i-m) components along three different planes, namely OX (with Y=Z=0 and X=$\pm5R_E$ in red), dawn (with Y=$+2R_L$ and Z=0 in blue), and dusk (with Y=$-2R_M$ and Z=0 in green). The plots are taken along OX at $\pm5R_E$ on both the day and night sides of the moon. The linear plots at dawn and dusk account for the asymmetry of the solar wind parameters.}}
	\label{emfvel}
\end{figure}
%%%%Figure 7
\begin{figure}[ht]
	\centering
	\includegraphics[width=0.9\linewidth]{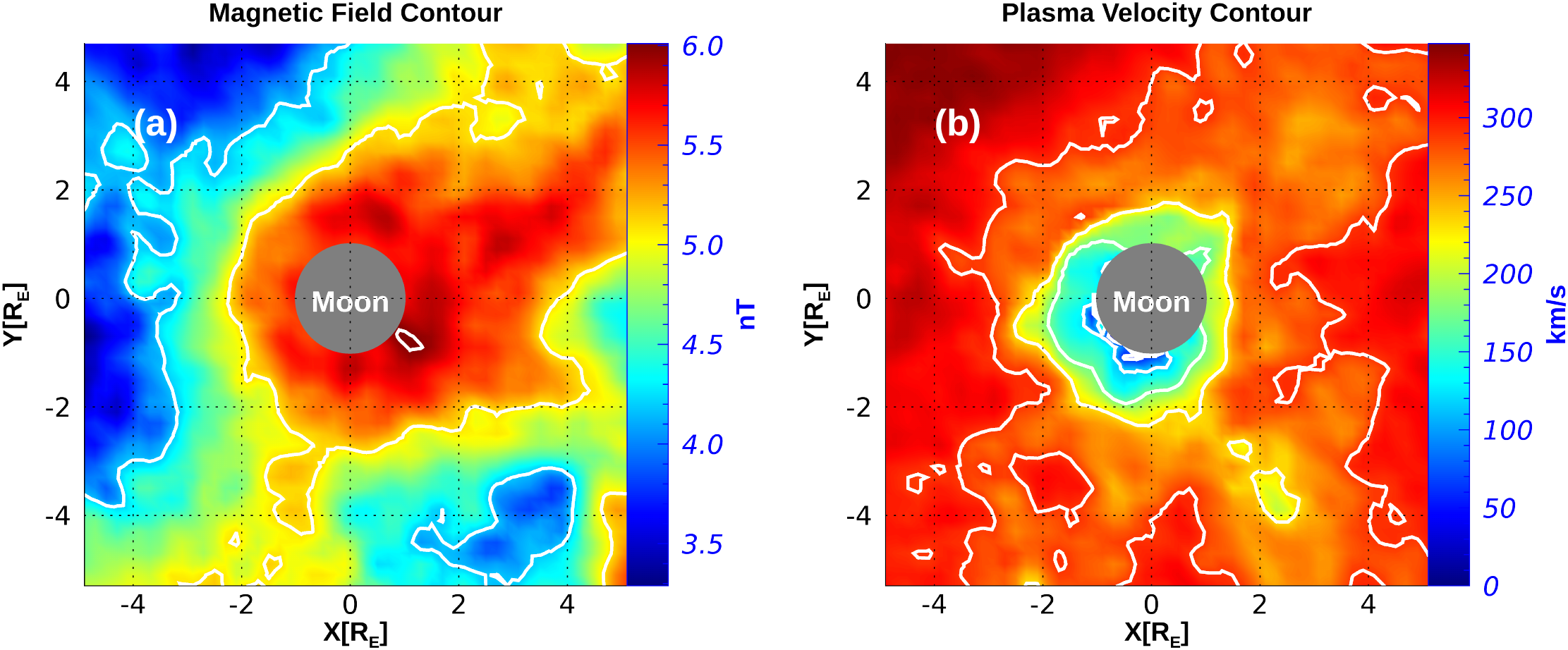}
	\caption{\fontfamily{cmtt} {This figure shows two contours for magnetic field and plasma velocity taken in equatorial plane around the moon at $\pm 5R_E$ in both directions. color codes are scaled to real values.  } }
	\label{xyfields}
\end{figure}
%%%%Figure 8
\begin{figure}[ht]
	\centering
	\resizebox{1.\textwidth}{!}{\begin{tabular}{cc}
			
			\includegraphics[width=0.9\linewidth]{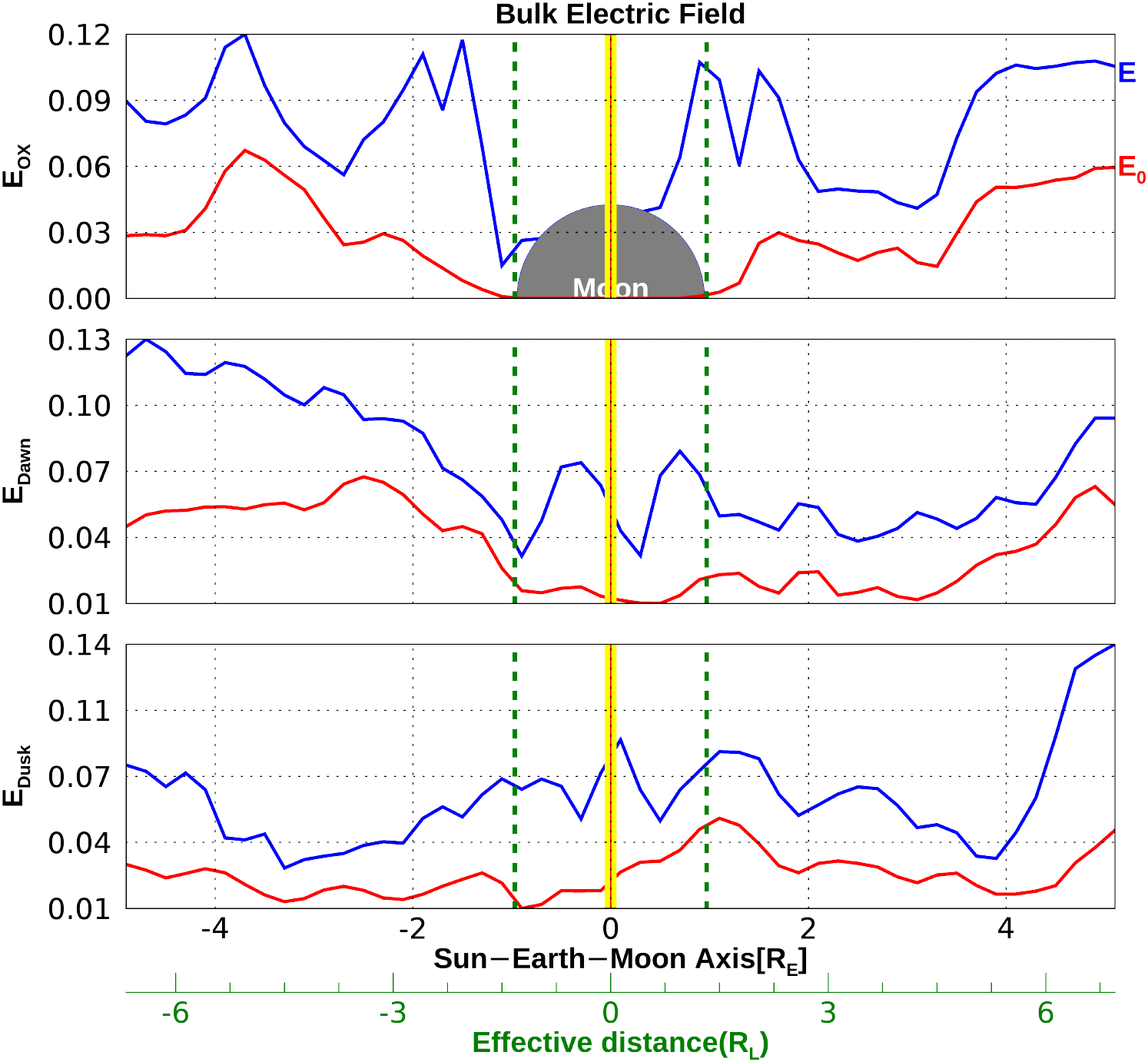}	& \includegraphics[width=0.9\linewidth]{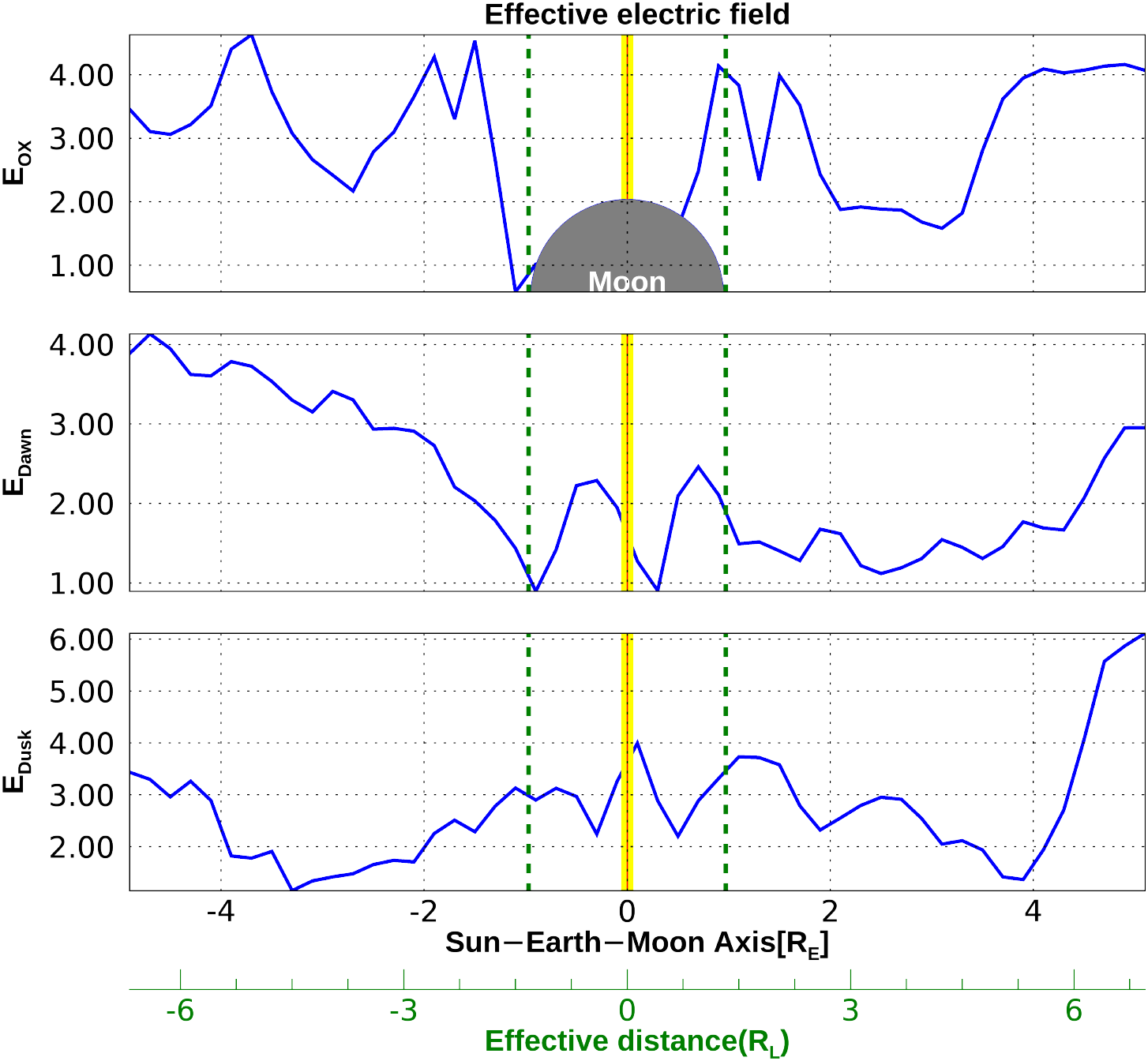}\\
			A&B\\ 
	\end{tabular}}
	\caption{\fontfamily{cmtt} This figure presents a zoomed in depiction of the moon, where the radius is illustrated as being five times greater than the ratio of the moon's radius to the Earth's radius (\(R_L/R_E\)) .The total electric field ($E.F.$) of the solar wind plasma is plotted in blue, and the induced electric field from charge separation ($E_0=m_e v_e \omega_{pe}/q_0$) is plotted in red. These fields are plotted along the OX direction at Y=0 and Y=$\pm 1R_L$ in the dawn and dusk regions, respectively. Panel-B shows the effective E.F. normalized over $E_0$ to represent the total E.F. in the vicinity of the lunar surface along the Sun-Earth-Moon Axis.}
	\label{evse0}
\end{figure}
%%%%Figure 9
\begin{figure}
	\centering
	\begin{tabular}{cc}
		\includegraphics[width=0.45\linewidth]{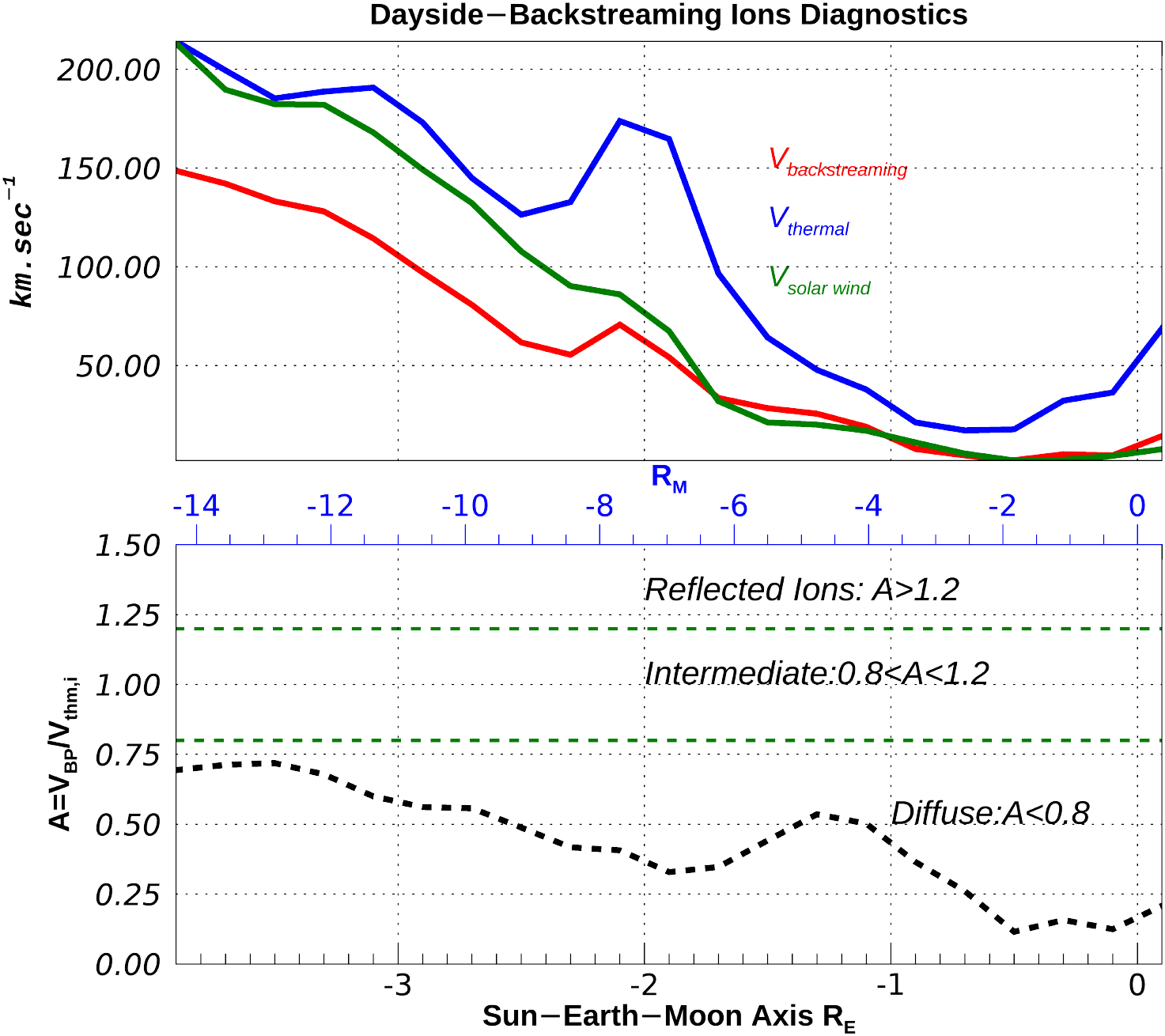}	& \includegraphics[width=0.45\linewidth]{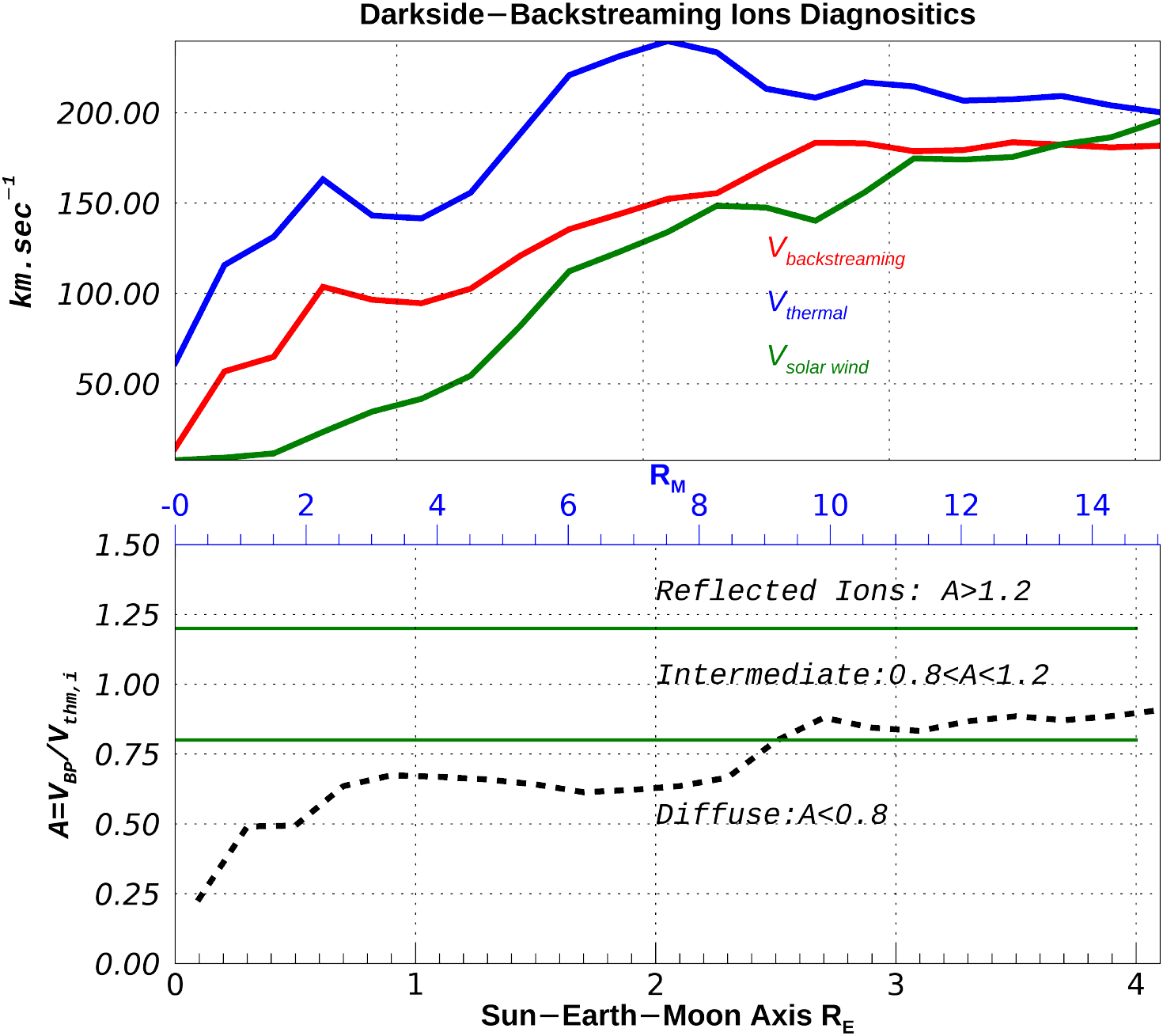} 	\\
	\end{tabular}
	\caption{\fontfamily{cmtt}\selectfont {Previous observations \cite{Poppe2014} and PIC simulations \cite{Deca2015} have revealed that some solar wind ions backstream against the solar wind inflow. This figure presents the characteristics of the backstreaming ions on both sides of the Moon along the Sun-Earth-Moon direction. The backstreaming ions are characterized based on their $\mathrm{A-factor}$, defined as the ratio of backscattered velocity to ion thermal velocity \cite{Bonifazi1981}. If $\mathrm{A}>1.2$, the net flow is a reflection; if $0.8<\mathrm{A}<1.2$, the flow is intermediate, and finally, if $\mathrm{A}<0.8$, the flow is diffuse. The comparison is made at both sides of the Moon at $\pm$4$R_E$ (approximately 15$R_M$). Most backstreaming ions are characterized as diffuse ions on both sides of the Moon.}}
	\label{backstreaming}
\end{figure}
% Figure 10
\begin{figure}[ht]
	\centering
	\includegraphics[width=0.75\linewidth]{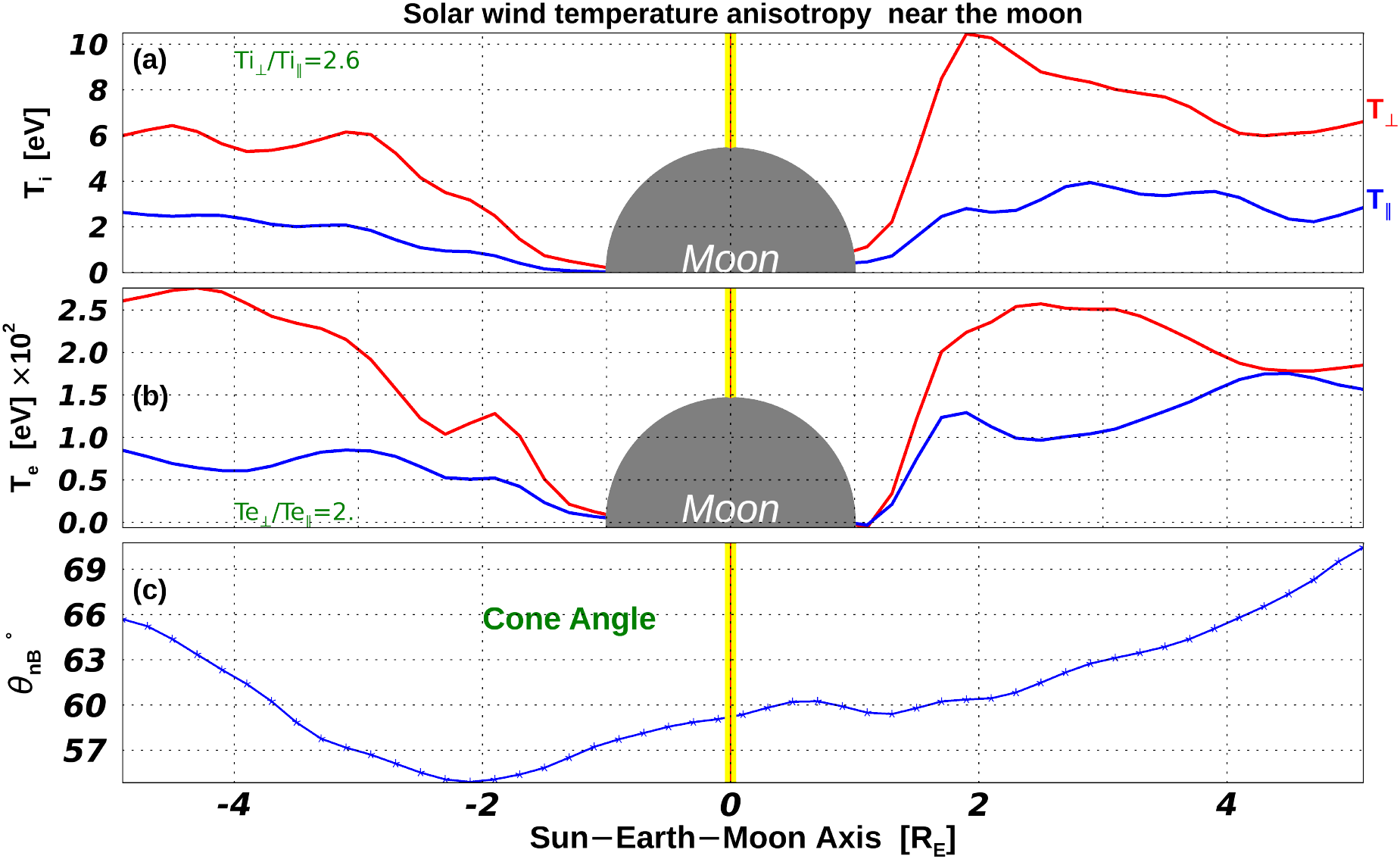}
    \caption{\fontfamily{cmtt}\selectfont {The figure depicts ion temperature anisotropy (a) and electron temperature anisotropy (b), along with cone angle \(\theta_{nB}\)\ ) represented in (c). The measurements were taken in the magnetotail region near the lunar position as shown in this figure at a distance of $\pm 5 R_E$. The data reveals that the ratio of perpendicular to parallel ion temperature anisotropy ($Ti_{\bot}/Ti_{\parallel}$) is 2.6, whereas the ratio of perpendicular to parallel electron temperature anisotropy ($Te_{\bot}/Te_{\parallel}$) is 2.1.}}
	\label{anisotemp}
\end{figure}
%%%%Figure 11
\begin{figure}[ht]
	\centering
	\includegraphics[width=1.0\linewidth]{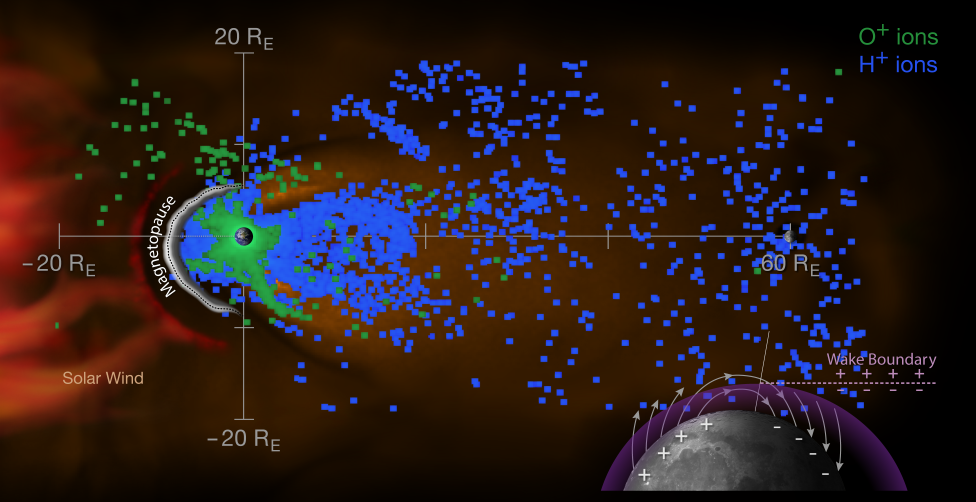}
    \caption{\fontfamily{cmtt}\selectfont {This figure succinctly summarizes our findings. It showcases the interaction between the solar wind (SW) and the Earth's magnetosphere (MS), leading to the derivation of the magnetopause (MP) shape (represented by the curved gray line). The figure clearly illustrates the dusk-dawn asymmetry. Furthermore, it depicts the coupling between the magnetosheath (MS) and the ionosphere (IS) through the escape of  \(\mathrm{H^+}\)  ions (blue) and  \(\mathrm{O^+}\)  ions (green), which are superimposed on the background color representing the plasma distribution of SW ions. Additionally, the figure demonstrates the charging of the lunar surface on both its day and night sides.}}
	\label{chart}
\end{figure}
\end{document}